\documentclass{article}
\usepackage{graphicx}
\usepackage{amsmath}
\usepackage{authblk}
\usepackage{url}
\usepackage{dsfont}
\usepackage{indentfirst}
\usepackage{amsfonts}

\usepackage{setspace}
\usepackage{mathptmx}
\usepackage[margin= 1.25 in]{geometry}

\usepackage[utf8]{inputenc}
\usepackage[english]{babel}
\newtheorem{theorem}{Theorem}

\newtheorem{definition}{Definition}[section]

\usepackage{xparse}
\makeindex

\let\oldsection\section  

\RenewDocumentCommand\section {s o m o}
   {
     \IfNoValueTF{#2}
          {                                                      
             \IfBooleanTF{#1}                       
                 { \oldsection*{#3} }                                
                 { \oldsection {#3} }                                
          }
           {                                                     
              \IfBooleanTF{#1}
                 { \oldsection*[#2]{#3} }
                 { \oldsection[#2]{#3}  }
         }   
    \IfNoValueF{#4}                
        {
            \label{sec:#4}
         }
  }

\DeclareUnicodeCharacter{2212}{-}

\begin{document} 

\begin{titlepage}
   \begin{center}
       \vspace*{1cm}
 
       \text{Rational Kernel on Pricing Models of Inflation Derivatives}
       
       \vspace{1cm}
      by \\
       \vspace{1cm}
 
       \text{Yue Zhou}
 
        \vspace{1cm}
A thesis submitted in partial fulfillment \\
       \vspace{1cm}
Of the requirements for the degree of \\
       \vspace{1cm}
Master of Science \\
       \vspace{1cm}
Courant Institute of Mathematical Sciences \\
       \vspace{1cm}
New York University \\
       \vspace{1cm}
September, 2019\\
 
       \vspace{1cm}
Under the Supervision of Professor Leon Tatevossian

   \end{center}
\end{titlepage}

\centerline{\textbf{Acknowledgements}}

I sincerely appreciate my advisor, Professor Leon Tatevossian, for his guidance and the greatest help on my academy, as well as on my life, especially his motivation and positivity on curing my depression, which possessed me for the past four years. Every time when I visit my advisor, I can feel his wisdom and confidence toward both academy and life. It is truly my honor to have this opportunity to learn from him. \\

I further wish to thank Professor S. R. Srinivasa Varadhan, for the guidance on my academy and his kindness to me. Without his encouragement, I would not get intrigued by mathematics. I sincerely appreciate Professor Varadhan for giving me this opportunity to take his real variable class. Moreover, I really appreciate Professor S. R. Srinivasa Varadhan for his time and effort on reading my thesis. \\

I gratefully thank Professor Bruce Hughes, who taught me topology during my undergraduate study. His thoughtfulness is a gift that I will always treasure. His charming personality definitely retains in my heart as the lighthouse leading my direction when my mood turns to be foggy. \\

I sincerely thank Professor Ralph McKenzie, for his guidance and kindness. I am offered many opportunities from him, on researches, conferences, and participation in his class. His profound thought has encouraged me on challenging myself all the time. Without him, I would not relish in pure mathematics as much as I do. \\

I sincerely thank my undergrad advisor, Professor Glenn Webb, for everything that he guides me along, he teaches me, and he helps me on. His enthusiasm toward academic researches allures me and has been my guidance. It is my great honor to be his student, having the opportunity to learn from him, to observe his persistence, time, and effort that he dedicates to both research and teaching.\\  

I appreciate thank Vanderbilt university for enriching my positive attitude toward life and academic research. I appreciate New York University for training me like training a warrior, who enjoy the process of persisting on challenging myself. \\

From the bottom of my heart, I appreciate my parents for the companion, the unlimited support and love through my life. I sincerely appreciate that every person who expects me to do well. I might have disappointed them and even myself, but I would never let myself down. \\

Anyway, what happened in the past remains in past, the attitude toward life retains: Citius, Altius, Fortius. I once thought academy was dry and fastidious, which I would think as being fruitful and fascinating. I once thought competition was for lacking of resources, which I would think as for training me as a qualified researcher. As time passing by, materialistic objects shall fade away, meanwhile, knowledge will shine out. Simply, I decide to utilize my time and effort on thrusting the boundary of human beings' acknowledge. Human beings stand as one. \\

Life is bittersweet, it's bitter at the beginning, then all the sweetness comes in. \\

The last but not the least, I love math for its own beauty.

\begin{abstract}
The aim of this thesis is to analyze and renovate few main-stream models on inflation derivatives. 
In the first chapter of the thesis, concepts of financial instruments and fundamental terms are introduced, such as coupon bond, inflation-indexed bond, swap.

In the second chapter of the thesis, classic models along the history of developing quantified interest rate models are introduced and analyzed. Moreover, the classification of interest rate models is introduced to help audiences understand the intrinsic ideology behind each type of models. 

In the third chapter of the thesis, the related mathematical knowledge is introduced. This part has the contribution on understanding the terms and relation among terms in each model introduced previously.

In the fourth part of the thesis, the renovation of HJM frame work is introduced and analysis has been initiated. 
\end{abstract}

\tableofcontents

\section{Financial Terminology}
\subsection{Bond}
\subsubsection{Definition}
In finance, a bond is an instrument of indebtedness of the bond issuer to the holders, under which the issuer owes the holders a debt and is obliged to pay them interest at fixed intervals(semiannual, annual, sometimes monthly) and to repay the principal at a later date, termed the maturity date.\cite{arthur2003economics} On the one hand, bonds are applied as one of the methods that government and corporations commonly use to borrow money from investors.  On the other hand, bonds are considered as one of the ways that investors make profits of. Bond can be thought as a form of loan or IOU: the holder of the bond is the lender (creditor), the issuer of the bond is the borrower (debtor), and the coupon is the interest. Bonds provide the borrower with external funds to finance long-term investments, or, in the case of government bonds, to finance current expenditure. Certificates of deposit (CDs) or short-term commercial paper are considered to be money market instruments and not bonds: the main difference is the length of the term of the instrument.\cite{arthur2003economics}Thus, bonds provide a solution for both sides to quench their financial need. Very often the bond is negotiable, that is, the ownership of the instrument can be transferred in the secondary market. This means that once the transfer agents at the bank medallion stamp the bond, it is highly liquid on the secondary market. It was first issued by Venetian government in thirteen century designed to pay 5\%\ interest rate twice a year and never mature. During the World War II the United States government raised almost two hundred billion U.S. dollars solely in war bonds and the money solidified the foundation of Allies’ victory. \cite{mobius2012bonds}

\subsubsection{The Functionality of Bond}
When governments or corporations urge to acquire money for long-term investments or to finance current expenditure in the case of government bond, they may issue bonds. Since the bonds are not only the contracts between issuers and buyers but also assets, most of the bonds can be traded publicly or over-the-counter(OTC) from one bond holder to another where over-the-counter(OTC) refers to the process of how securities are traded for the companies that are not listed on a formal exchange. Such securities are traded via a broker-dealer network as opposed to on a centralized exchange, since the securities do not fulfill the requirements of being listed on a standard market exchange. The bond is asset due to its par value (face value) paid at maturity and its interest rate paid periodically at certain intervals. To be addressed, some bonds do not have interest rate, which will be introduced later. In addition, it is also common for a bond issuer to repurchase bonds from buyers if the interest rate decline or the credit of the issuer has improved, which allows the issuer to reissue new bonds at a lower cost. \cite{bonds}

\subsubsection{Characteristics of Bond}
There are few concepts to be introduced in this paragraph, such as face value, coupon rate, coupon date, the maturity date, the issue price, the market value. 
\begin{enumerate}
\item Face value 

This is the dollar value of a bond that is to be repaid at maturity. In other words, this is what the investor has loaned to the issuer and will get back. Generally, bonds are issued in denominations of US \$1,000, but some can be in denominations of US \$5,000 or even larger. \cite{mobius2012bonds}

\item Coupon rate

Coupon rate is the amount of interest paying on quarterly, semiannual, or annual basis in a percentage term. The coupon rate retains in the entire life span of a bond. Some bonds, such as nominal bonds, have a fixed interest rate at each interval of payment. On the contrary, the other bonds, such as inflation-indexed bonds, have a floating rate of interest, where the rate is adjusted periodically in line with some measures of market rates. The floating interest rate only correlates to the interest-rate conditions, and has no correlation with the issuer’s creditworthiness. In the case of inflation-indexed bonds, they are issued by U.S. department of treasury. The Coupon rate of a bond as compared to the interest rates in the economy determines whether a bond will trade at par, below par, or above its par value.

\item Coupon date

One of the dates on which bondholders are sent coupon payments. That is, the coupon dates are the dates on which bondholders receive the interest that they are guaranteed. Coupon dates are fixed for bonds, and usually occur twice a year.

\item Maturity date

In finance, maturity or maturity date refers to the final payment date of a loan or other financial instrument, at which point the principal (and all remaining interest) is due to be paid.

\item Issue price

The issue price of shares is the price at which they are offered for sale when they first become available to the public.

\item Market value

Market value is the price an asset would fetch in the marketplace. Market value is also commonly used to refer to the market capitalization of a publicly traded company, and is obtained by multiplying the number of its outstanding shares by the current share price. Market value is easiest to determine for exchange-traded instruments such as stocks and futures, since their market prices are widely disseminated and easily available, but is a little more challenging to ascertain for over-the-counter instruments like fixed income securities. However, the greatest difficulty in determining market value lies in estimating the value of illiquid assets like real estate and businesses, which may necessitate the use of real estate appraisers and business valuation experts respectively.\cite{market}

\end{enumerate}

\subsubsection{Zero-Coupon Bond}
One of the special cases is zero-coupon bond. A zero-coupon bond is a bond where the face value is repaid at the time of maturity. \cite{zerocoupon}Zero-coupon bonds do not have a periodic interest payment. Instead, zero-coupon bonds are sold at deep discounts to their face value. For example, the United States department of treasury sold 26-week bills with \$100 face value for the price \$98.78 on October 18th, 2018. The gap between its sold value and face value reveals a total annual yield of 2.479\%. Since zero-coupon bonds do not hold a coupon rate and its face value is fixed, we can calculate and estimate the “interest rate” of a coupon bond. Such “interest” is defined as imputed interest, which is an estimated interest rate for the bond and not an established interest rate. Though there is no transaction of coupon payment before the coupon bond matures, bond holders still have the obligation to pay federal, state, and local income taxes based on the impute interest. Thus, the imputed interest on the bond is subject to income tax.

\subsection{Inflation}
\subsubsection{Definition}
Inflation is a quantitative measure of the rate at which the average price level of a basket or selected goods and services in an economy arising over a period of time. In another word, inflation indicates an erosion of purchasing power in an unit currency.\cite{inflation}

There is a controversial problem of defining inflation, as well as the relations among products and services: the relative value of any two products/services changes with time. This relation indicates that the inflation, "value of money", is subjective to some extent. This is the reason why one often refers to a basket or a scale of goods and services to determine the level of inflation. 

\subsubsection{Importance}
Inflation affects economies in various positive and negative ways. The negative effects of inflation include an increase in the opportunity cost of holding money, uncertainty over future inflation which may discourage investment and savings, and if inflation were rapid enough, shortages of goods as consumers begin hoarding out of concern that prices will increase in the future. Positive effects include reducing unemployment due to nominal wage rigidity,\cite{importance}allowing the central bank more leeway in carrying out monetary policy, encouraging loans and investment instead of money hoarding, and avoiding the inefficiencies associated with deflation.
\subsubsection{Inflation Rate}
The inflation rate measures the percentage change in purchasing power of a currency which has a positive correlation with the cost of price and a negative correlation with the purchasing power. 
\subsection{Price Index}
\subsubsection{Definition}
Price index is a normalized average (typically a weighted average) of price relatives for a given class of goods or services in a given region, during a given interval of time. It is a statistic designed to help to compare how these price relatives, taken as a whole, differ between time periods or geographical locations.\cite{price}
\subsubsection{Importance}
The price index is not only meaningful toward the calculation of inflation, more importantly, is used widely in informing government economic policy, monetary policy, wage bargaining, the level of pension increases etc.

\subsubsection{How to calculate the price index}
Given a set $C$ of goods and services, the total market value of transactions in $C$ in some period $t$ would be
\begin{displaymath}
\sum_{c \in C}(p_{c,t} \cdot q_{c,t})
\qquad
\end{displaymath}
where

$p_{c,t}$ represents the prevailing price of $c$ in period $t$

$q_{c,t}$ represents the quantity of $c$ sold in period $t$

\noindent If, across two periods $t_{0}$ and $t_{n}$, the same quantities of each good or service were sold, but under different prices, then
\begin{displaymath}
q_{c,t_n}=q_c=q_{c,t_0} 
\qquad 
\forall c \in C
\qquad
\end{displaymath}
and
\begin{displaymath}
P=\frac{\sum_{c \in C}(p_{c,t_n} \cdot q_c)}{\sum_{c \in C}(p_{c,t_0} \cdot q_c)}
\qquad
\end{displaymath}
would be a reasonable measure of the price of the set in one period relative to that in the other, and would provide an index measuring relative prices overall, weighted by quantities sold. Of course, for any practical purpose, quantities purchased are rarely if ever identical across any two periods. As such, this is not a very practical index formula.

One might be tempted to modify the formula slightly to
\begin{displaymath}
P=\frac{\sum_{c \in C}(p_{c,t_n} \cdot q_{c,t_n})}{\sum_{c \in C}(p_{c,t_0} \cdot q_{c,t_0})}
\qquad
\end{displaymath}

\subsubsection{Criteria of constructing a price Index}
The criteria of constructing a price index differs from country to country. 
For example, the Bureau of Labor Statistics(BLS) publishes the index called Consumer Price Index(CPI): 
CPI reflects spending patterns for each of two population groups: all urban consumers and urban wage earners and clerical workers. The all urban consumer group represents about 93 percent of the total U.S. population. It is based on the expenditures of almost all residents of urban or metropolitan areas, including professionals, the self-employed, the unemployed, and retired people, as well as urban wage earners and clerical workers. Not included in the CPI are the spending patterns of people living in rural non-metropolitan areas, those in farm households, people in the Armed Forces, and those in institutions, such as prisons and mental hospitals. Consumer inflation for all urban consumers is measured by two indexes, namely, the Consumer Price Index for All Urban Consumers (CPI-U) and the Chained Consumer Price Index for All Urban Consumers (C-CPI-U).
The Consumer Price Index for Urban Wage Earners and Clerical Workers (CPI-W) is based on the expenditures of households included in the CPI-U definition that also meet two additional requirements: more than one-half of the household's income must come from clerical or wage occupations, and at least one of the household's earners must have been employed for at least 37 weeks during the previous 12 months. The CPI-W population represents about 29 percent of the total U.S. population and is a subset of the CPI-U population.
Similarly, in the UK the Office for National Statistics(ONS) measures the retail price index(RPI). Meanwhile, the ONS also calculates the Consumer price index(CPI) which mainly differs from CPI by excluding the housing-related expenses such as mortgage payments,council tax. In France, the INstitut National de la Statistique et des Etudes Economiques(INSEE) publishes the consumer price index for all France(FCPI) for French inflation-linked bonds. Moreover, the 
Thus, each country or region has its unique method on calculating the price index by constructing the index formula based on its unique category of goods and services based on a unique weighted scale.
\subsection{Consumer Price Indices (CPI)}
\subsubsection{Definition}
The CPI is a statistical estimate constructed using the prices of a sample of representative items whose prices are collected periodically and it is used to measures changes in the price level of market basket of consumer goods and services purchased by households.

In fact, different countries and different regions utilize distinct indices to measure the inflation in their own countries and regions. For example, the government of US uses CPI (Consumer Price Index) to measure the inflation in USA and the government of United Kingdom uses RPI (Retail Price Index) to measure the inflation.

\subsubsection{Explanation of CPI}
CPI reflects spending patterns for each of two population groups: all urban consumers and urban wage earners and clerical workers. The all urban consumer group represents about 93 percent of the total U.S. population. It is based on the expenditures of almost all residents of urban or metropolitan areas, including professionals, the self-employed, the unemployed, and retired people, as well as urban wage earners and clerical workers. Not included in the CPI are the spending patterns of people living in rural non-metropolitan areas, those in farm households, people in the Armed Forces, and those in institutions, such as prisons and mental hospitals. Consumer inflation for all urban consumers is measured by two indexes, namely, the Consumer Price Index for All Urban Consumers (CPI-U) and the Chained Consumer Price Index for All Urban Consumers (C-CPI-U).
The Consumer Price Index for Urban Wage Earners and Clerical Workers (CPI-W) is based on the expenditures of households included in the CPI-U definition that also meet two additional requirements: more than one-half of the household's income must come from clerical or wage occupations, and at least one of the household's earners must have been employed for at least 37 weeks during the previous 12 months. The CPI-W population represents about 29 percent of the total U.S. population and is a subset of the CPI-U population.

\subsubsection{Relation between CPI and inflation rate}
The inflation rate formula uses the Consumer Price Index announced by the Bureau of Labor Statistics in the United States. Meanwhile, some other similar indices may be applied to the inflation rate formula. If another index is applied, the terms of CPI in the inflation rate formula are substituted by the alternative index. 

The subscript "i" indicates the initial consumer price index at the time interval i that the formula intakes. As such, the subscript "j" indicates the ending consumer price index at the time interval j.

\begin{displaymath}
Inflation \ Rate = \frac{CPI_j-CPI_i}{CPI_i}
\qquad
\end{displaymath}
where

$i$ indicates the initial consumer price index at the time interval $i$ 

$j$ indicates the ending consumer price index at the time interval $j$

In UK's economic system, the calculation of year-on-year inflation rate has a different method and is based on a different index, compared to US economic system.

\subsection{Inflation-Indexed Bond (IIB)}
\subsubsection{Definiton}
Daily inflation-indexed bonds (also known as inflation-linked bonds or colloquially as linkers) are bonds where the principal is indexed to inflation or deflation on a daily basis. They are thus designed to hedge the inflation risk of a bond. \cite{shiller2003invention}The inflation-linked market primarily consists of sovereign bonds, with privately issued inflation-linked bonds constituting a small portion of the market.\cite{campbell2009understanding}

\subsubsection{History}
The earliest recorded inflation-indexed bonds were issued by the Commonwealth of Massachusetts in 1780 during the Revolutionary War.\cite{IIB} Much later, emerging market countries began issuing ILBs in the 1960s. In the 1980s, the UK was the first major developed market to introduce “linkers” to the market. Several other countries followed, including Australia, Canada, Mexico and Sweden. In January 1997, the U.S. began issuing Treasury Inflation-Protected Securities (TIPS), now the largest component of the global ILB market.\cite{seach2} Today inflation-linked bonds are typically sold by governments in an effort to reduce borrowing costs and broaden their investor base. Corporations have occasionally issued inflation-linked bonds for the same reasons, but the total amount has been relatively small.\cite{article1}

\subsubsection{The purpose of IIB}
Though the main reason of supplying inflation-indexed bonds to investors is to preserve a certain real return and the main reason of offering inflation-indexed bonds from issuers is to obtain a real debt burden, however, inflation-indexed bonds cannot secure complete real value certainty due to the three reasons following.

\begin{enumerate}
\item Every index has its own limitation on providing the precise consumption basket to individual investors. For example, different market baskets (or commodity bundles) has their own numerical proportions on a different collection of selected price categories. (One of the ubiquitous market baskets is the basket of consumer goods, which conducts the Consumer Price Index(CPI). Specifically, the categories included in the basket are food and beverages, housing, apparel, transportation, medical care, recreation, education and communication, other goods are services. )Only if the proportional collection of goods and services from given index has the complete match with investors’ need of consuming, otherwise the index cannot provide the precise consumption basket to investors. \cite{seach3}
There exists a time difference between when the price index value is calculated and when the price index value is announced. Thus, by the time when the price index is published, the published price index applies to the price index where the collections of goods and services have been collected. Though this time lagging does not cause much effect in a country where the inflation rate is stable, it causes influences in the countries with highly variable price indices. In fact, the calculation of price index takes from days to over a year, which indicates the essentialness of the concern. 

\item The last reason is due to the complexity of the system of tax. It is citizens free choice to decide how they split their income as pre-tax or post-tax. In the case of pre-tax, citizens pay income tax as usual; however, in the case of post-tax, citizens do not pay tax but only can withdraw the money from their bank accounts when they are retired. Thus, the fact of existence of pre-tax and post-tax causes an instability of the cash flow influencing the inflation. Meanwhile, the uncertainty of that tax regimes maintain or renovate the tax policy as the same during the life time of an indexed bond evokes a third factor that addresses the concern when modelling the inflation-indexed bond.
\end{enumerate}

\subsubsection{Characteristics}
There are three influential factors of inflation-indexed bond yields:
\begin{enumerate}
\item Current and expected future short-term real interest rates
\item Differences in expected returns on long-term and short-term real bonds caused by risk premia (which can be negative if inflation-indexed bonds are valuable hedges)
\item Differences in expected returns on long-term and short-term bonds caused by liquidity premia or technical factors that segment the bond market.
\end{enumerate}

The priority of considering the three factors above in our model is hierarchy of the number from first to third.
The generic hypothesis of the term structure indicates that the only time-varying factor is the current and expected future short-term real interest rates; while the other two are constant. On the contrary, in the case of constructing models of nominal treasury bonds, it is essential to consider the possibility that risk and liquidity premia are time-varying.\cite{seach2}

Daily inflation-indexed bonds pay a periodic coupon that is equal to the product of the principal and the nominal coupon rate.

For some bonds, such as in the case of TIPS, the underlying principal of the bond changes, which results in a higher interest payment when multiplied by the same rate. For example, if the annual coupon of the bond were 5\%\ and the underlying principal of the bond were 100 units, the annual payment would be 5 units. If the inflation index increased by 10\%\ ,the principal of the bond would increase to 110 units. The coupon rate would remain at 5\%\ , resulting in an interest payment of 110 x 5\%\ = 5.5 units.

For other bonds, such as the Series I United States Savings Bonds, the interest rate is adjusted according to inflation.\cite{bonds}

The relationship between coupon payments, break-even daily inflation and real interest rates is given by the Fisher equation. A rise in coupon payments is a result of an increase in inflation expectations, real rates, or both.\cite{seach2}

\subsubsection{Index of IIB}

In general, indexed bond has the property of which the cash flows are linked to a specific price index; especially to the movement of the price index. The major price index has been broadly used by different countries is the domestic Consumer Price Index (CPI) of their own countries, since CPI reflects all income and principal payments adjusted for inflation. Meanwhile, some other types of inflation indices are occasionally used, such as wholesale prices, average earnings and the GDP deflator. (page 6 of the book: inflation – indexed securities: bonds, swaps and other derivatives.)
According to Deacon (Inflation-Indexed Securities: Bonds, Swaps and Other Derivatives, P74), In reality, index-linked bonds are not perfectly indexed to inflation. Most fundamentally, the index used to inflate the cash flows may not be an accurate measure of “true” inflation. Most index-linked bonds currently in existence are linked to a consumer price index(CPI). As discussed in Chapter 3, for many classes of investors consumer price inflation is unlikely to be the most relevant measure – pension funds, for example, often have liabilities linked to wage growth (in addition to those linked to consumer prices), and so only a bond indexed to a measure of wage inflation will provide them with a real bond that truly allows them to hedge such liabilities. In this case the inflation terms in Equation (5.2) cannot cancel one another because the measure used in the numerator to scale up cash flows will not be the same as the one used by investors in the denominator to discount them. Even for those investors for whom the Consumer Price Index (CPI) is the relevant one, there is often debate over how this index is best constructed. For example, monetary authorities in the UK monitor at least four measures: the standard Retail Prices Index (RPI), the Retail Prices Index including mortgage, and the Harmonised Index of Consumer Prices (HICP). Whereas the nominal cash flows of index-linked gilts are defined using the RPI measure, the government’s inflation target is currently defined with reference to RPIX. If RPIX is viewed to be the “true” measure of consumer price inflation in the UK, then index-linked gilts will be imperfect real bonds to the extent that RPI differs from RPIX.

\subsubsection{The value of IIB}
The payment of inflation-indexed bond has reflection on fixed real value (inflation adjusted), which has its correspondence of currency’s buying power at mature time. On the contrary, the payment of nominal bond has fixed amount on payment at maturity. (Thus, in order to simulate the price of inflation-indexed bond in mathematical models, there are additional volatility factors involved than the traditional models for nominal bonds. ) Specifically, the inherent value of indexed bond can be summarized and decomposed as two factors: 
\begin{enumerate}
    \item The real rate of return
    \item The compensation of seizing the purchasing power caused by inflation
\end{enumerate}

The real rate of return is predictable and calculated directly by applying the seasonal return rate: accumulating the seasonal return rate season by season until the bond matures. Meanwhile, the nominal return at maturity is uncertain because the nominal return depends on the future movement path of price index.  

\subsubsection{Real yield}
The real yield of any bond is the annualized growth rate, excluding the rate of inflation over the same period. This calculation is often difficult in principle in the case of a nominal bond, because the yields of such a bond are specified for future periods in nominal terms, while the inflation over the period is an unknown rate at the time of the calculation. However, in the case of inflation-indexed bonds such as TIPS, the bond yield is specified as a rate in excess of inflation, so the real yield can be easily calculated using a standard bond calculation formula.

\subsubsection{Calculating the yield of Inflation Coupon Bond}
The nominal interest rate is not the "true return" that an investor gains from the inflation-indexed bond. Meanwhile, the "true return" relates to the purchasing power of a currency. For example, if the inflation happens, the return in currency that investor obtained from the bond devalues, as well as the face value of the bond. Thus, the real return rate gained by investor are calculated by eliminating the inflation rate from the nominal interest rate.\cite{seach5}Furthermore, the calculation of real rate is calculated as:\\
$r = \frac{1+n}{1+i}-1$ where r represents the real rate, n represents the nominal rate, and i represents the inflation rate.\\
In addition, when the inflation rate turns to remain in a low rate, the relation between real rate and nominal rate becomes: $r \approx n- i$. \\
Indeed, this is Fisher equation:\\
$(1+r_r)(1+i)=(1+r_n)$\\
$r_r=r_n - i$\\
where $r_r$ is real rate, $r_n$ is nominal rate, and $i$ is inflation rate.\\

\subsubsection{Primary Inflation-linked Instruments}
\begin{enumerate}
\item Year-on-Year Inflation-Indexed Swap(YYIIS)

The Year-on-Year Inflation-Indexed Swap(YYIIS) is a standard derivative product over Inflation rate. The underlying is a single Consumer Price Index (CPI).
The reason why it is called swap is that each year there is a swap of a fixed amount against a floating amount. However, in reality, there is only one way payment transacted: fixed amount - floating amount.\cite{seach8}

\item Zero-Coupon Inflation-Indexed Swap(ZCIIS)

The Zero-Coupon Inflation Swap (ZCIS) is a standard derivative product which payoff depends on the Inflation rate realized over a given period of time. The underlying asset is a single Consumer price index (CPI).
It is called Zero-Coupon because there is only one cash flow at the maturity of the swap, without any intermediate coupon.
It is called Swap because at maturity date, one counter-party pays a fixed amount to the other in exchange for a floating amount (in this case linked to inflation). The final cash flow will therefore consist of the difference between the fixed amount and the value of the floating amount at expiry of the swap.\cite{zeroswap}

\item What is the market and who trades

A market is one of the many varieties of systems, institutions, procedures, social relations and infrastructures whereby parties engage in exchange. In mainstream economics, the concept of a market is any structure that allows buyers and sellers to exchange any type of goods, services and information. The exchange of goods or services, with or without money, is a transaction.[1] Market participants consist of all the buyers and sellers of a good who influence its price, which is a major topic of study of economics and has given rise to several theories and models concerning the basic market forces of supply and demand. \cite{markets}

\item Inflation swap

An inflation swap is a contract used to transfer inflation risk from one party to another through an exchange of fixed cash flows. In an inflation swap, one party pays a fixed rate cash flow on a notional principal amount while the other party pays a floating rate linked to an inflation index, such as the Consumer Price Index (CPI). The party paying the floating rate pays the inflation adjusted rate multiplied by the notional principal amount. Usually, the principal does not change hands. Each cash flow comprises one leg of the swap.\cite{inflationswap}

\end{enumerate}

\subsubsection{Risk}
As with other investments, the price of IIBs can fluctuate, and if real yields rise, the market value of an IIB will fall. Real yields can rise due to an increase in inflation, without a corresponding increase in nominal yields. If held to maturity however, the market value fluctuations are irrelevant and an investor receives the par amount. In theory, a period of deflation could reduce this par amount. However, in practice most IIBs are issued with a deflation floor to mitigate this risk.\cite{seach2}

\subsection{Hedge with Inflation-Iinked Bond}
Inflation can have a dampening effect on fixed-income investments, reducing their purchasing power and cutting deeply into the real value of returns over time. This happens even if the inflation rate is relatively low. For example, if you have a portfolio that returns 9\%\, but the inflation rate is 3\%\, the true value of your returns has been cut by about a third. Inflation-linked bonds (IIBs), however, can help to offset (or hedge) that risk because they increase in value during inflationary periods.\cite{seach6}

\subsection{US treasury}
Treasury Inflation-Protected Security (TIPS) is a Treasury bond that is indexed to inflation to protect investors from the negative effects of rising prices. The principal value of TIPS rises as inflation rises. Inflation is the pace at which prices increase throughout the U.S. economy as measured by the Consumer Price Index or CPI.\cite{seach7} 

The United States Department of Treasury has started issuing Treasury Inflation-Protected Securities(TIPS), of which the price is linked to CPI-U, since 1997. The is currently over 300 billion USD outstanding with maturities up to thirty years. The uniqueness of US's TIPS is 
The most common derivatives are zero coupon bond and YoY swap.\cite{belgrade2004market}

\subsection{Inflation Derivatives}
\subsubsection{Definition}
Inflation derivatives are a subclass of derivative used by investors to manage the potential negative impact of rising inflation levels. It is announced by entities, such as governments, financial institutions and corporations, entities. Like other derivatives including options or futures, inflation derivatives allow individuals to participate in price movements of a market or index, in this case, a Consumer Price Index, a measurement of the general cost of goods and services in an economy. Doing so can help investors hedge against the risk of increasing prices eroding the real value of their investment portfolio.\cite{derivative}

Inflation-indexed derivatives are financial products announced by entities, such as governments, financial institutions and corporations, entities. The purpose of issuing inflation-indexed derivatives is to regulate the risks associated with variable rates of inflation.\cite{mercurio2005pricing}

\subsubsection{Nominal interest rate swap vs Real rate swap}
Nominal interest rate refers to the interest rate before taking inflation into account. Real rate swap is equal to nominal interest swap rate minus the corresponding inflation rate swap. As for modelling, the trend has been either to provide:
\begin{enumerate}
    \item 
    A model describing the relationship among: nominal rate, real rate, and the inflation rate, which is represented as the exchange rate between nominal rate and real rate.
    
    \item
    A market model that represents the inflation and applies the similar ideology as the representation of inflation returns in Brace Gatarek model. The first type of model alone these lines has been developed by Blegrade, Benhamou, Koehler. The advanced version of models in this type has been developed by Fabio Mercurio and Nicola Moreni.
    
\end{enumerate}

\subsubsection{Forward rate vs Expected spot rate}
The forward rate is a negotiated rate, which will be applied in the future, between two parties on a certain derivative. The forward rate applies to both parties to trade at this rate at a certain time in the future regardless what the exchange rate will be. In a short word, it is a rate on contract agreed by two parties.\cite{forwardrate} The expected spot rate is the rate of a financial instrument that on a future given date. Thus, the spot rate is the rate admitted by the entire market rather than a negotiated rate in a contract form.\cite{CORNELL197755}

\subsubsection{Discount Rate}
There are two different categories of defining the term of discount rate. The first definition of discount rate is the rate that is charged to financial institutions when they borrow money from the Federal Reserve. The second definition of discount rate is that the converting rate from a cash flow in the future to the cash flow in the present. In another word, it is the compensated rate from future value(maturity in most cases) to present value of a bond. Only the second definition is applied in this thesis. In the process of converting the value at maturity to the value at present, the major factors to be concerned are time value of money and uncertainty risk, which are analyzed in this thesis.

\subsubsection{Lag}
Regardless what the instrument is, the term lag refers to the delay of time. To be specific with inflation-indexed bond, there exists a lag of few months in the indexation. Due to the consumption of time to collect data and calculation for an index, the lag will at least one month out of date. If a cash flow requires to be fixed in advance, the lag can be even greater.(Cash flow, also known as cash flow loan, is a format of loan that requires time to be arranged.)

\subsubsection{Swap}
Swap is the arrangement of an exchange of two securities, interest rates, or currencies. Swap happens between two entities. The purpose of swap is to secure the mutual benefit of the both exchange parties. 

\subsubsection{Notional Principle amount}
The notional principal amount, in an interest rate swap, is the predetermined dollar amounts, or principal, on which the exchanged interest payments are based. The notional principal never changes hands in the transaction, which is why it is considered notional, or theoretical. Neither party pays nor receives the notional principal amount at any time; only interest rate payments change hands.

\subsubsection{Benchmark}
A benchmark is a standard against which the performance of a security, mutual fund or investment manager can be measured. Generally, broad market and market-segment stock and bond indexes are used for this purpose.

\subsubsection{London Interbank Offered Rate (LIBOR)}
The London Interbank Offered Rate (LIBOR) is a benchmark interest rate at which major global banks lend to one another in the international interbank market for short-term loans. It indicates borrowing costs between banks and is calculated and published each day by the Intercontinental Exchange (ICE).

\subsubsection{Leg}
A leg is one component of a derivatives trading strategy in which a trader combines multiple options contracts, futures contracts or—in rare cases—combinations of both to hedge a position, benefit from arbitrage or profit from a spread. Within these strategies, each derivative contract or position in the underlying security is called a leg. The cash flows exchanged in a swap are also referred to as legs.

\subsubsection{Breakeven Inflation rates}
The breakeven inflation rate is a market-based measure of expected inflation. It is the difference between the yield of a nominal bond and an inflation-linked bond of the same maturity.

\section{Model}
\subsection{Stochastic Discount Factor(Pricing Kernel)}
The term kernel is a common mathematical term used to represent an operator. The pricing kernel, as well as stochastic discount factor, is generally used in pricing the adjustment of the risk.

\section{the construction of the model}
\subsection{Set-ups}
As inflation is introduced earlier in the paper, Inflation is a change in price as measured by a price index, which defined in term of a basket of products and services. Thus, price index is one of the factors to influence inflation. 
The money-based economy which includes inflation is the nominal economy. On the contrary, the real economy is defined as inflation-free. 
\subsection{Model developed by Manning and Jones}
The methodology of this model is boot-strapping. 
The model of inflation derivatives by Manning and Jones has the following conditions:\\
1.Nominal price: $\mathbb{V}_n$ \\
2.Real price: $\mathbb{V_r}$ \\
3.Index value: $I$ \\
4.Price of nominal zero-coupon bond at time t with maturity T: $P_n(t,T)$ \\
5.Price of real zero-coupon bond at time t with maturity T: $P_r(t,T)$ \\
6.Determined forward index value at T from t: $I(t;T)$ \\
7.At time t enter into contract $I(t;T)$, short(sell) real ZCBs and long nominal ZCBs(buy): $I(t,T)=\frac{I(t)P_r(t,T)}{P_n(t,T)}$ \\
8.Forward inflation rate: $F^i(t;T_1,T_2) = \frac{I(t;T_2)}{I(t;T_1)} - 1 $\\
9.k=n,r, for the nominal and real forward rates: $F^k(t;T_1,T_2)=\frac{P_k(t,T_1)}{P_k(t,T_2)} - 1 $\\
10. Combine (8) and (9), we have: $1+F^n(t;T_1,T_2)=(1+F^r(t;T_1,T_2))*(1+F^i(t;T_1,T_2))$
Essentially, in this form the nominal rates include real rate and nominal rate; however, the nominal rates also include a third component,a risk premium. In this model, the risk premium term has been absorbed into the inflation term.(The risk premium term exists due to the unpredictability of inflation rate before the maturity.)\\
Since the nominal curve has been already simulated, the next step is either to produce a curve of forward index values and infer the real-rate term structure from it, or to produce a real-rate curve and imply the forward index values. The names of the methods are index projection(or inflation compensation), and real-rate methods, respectively. Meanwhile, the two methods are only the matters of approaching to the same solution.\\ 
11.The real dirty price per unit of real principle of a bond at time t:$B_r(t)=\sum^n_{i=1}(c+\delta_{in}P_r(t,T_i))$ where c is the real coupon and the remaining coupons to be paid at times $T_1...T_n$.
12. The nominal price generated by real price: $B_n(t) = (\frac{I_{ref}(t)}{I^{Base}_{ref}})B_r(t)$, where $I^{Base}_{ref}$ is the base value of the reference index for the bond. 
13. Inflation reference at day(d): $I_{ref} = CPI_{m-3}+ \frac{(nbd -1)}{ND_m} * [CPI_{m-2}-CPI_{m-3}]$\\
$CPI_{m-2}$: price index of $month_{m-2}$\\
$CPI_{m-3}$: price index of $month_{m-3}$\\
nbd: actual number of days since the start of the month\\
NDm: number of days in $month_m$\\
For example, if the given index reference date is July 25, 2006. \\
The inflation reference = $CPI(Apr06)+ \frac{25th -1}{31} *[CPI(May06) - CPI(Apr06)]$\\
14.The nominal price: $B_n(t)$ = $\sum^n_{i=1}\frac{I_{ref}(t;T_i)}{I^{Base}{ref}}(c+\sigma_{in})P_n(t,T_i)$\\
15.$dI(t) = I(t)[\mu_I(t) \ dt+\overrightarrow{\sigma}(t) \ d\overrightarrow{W}(t)]$\\
16.$df_n(t,T) = \alpha_n(t,T) \ dt + \overrightarrow{\sigma}_n(t,T) \ d\overrightarrow{W}(t)$ \\
17.$df_r(t,T) = \alpha_r(t,T) \ dt + \overrightarrow{\sigma}_n(t,T) \ d\overrightarrow{W}(t)$\\
18. The dynamics of the ZCBs: $d(logP_k(t,T))=[r_k(t)-\int^T_t\alpha_k(t,u) \ du] \ dt + \overrightarrow{\sum}_k(t,T) \ d\overrightarrow{W}(t),k = n,r$\\
19.ZCB(Zero Coupon Bond) volatility: $\overrightarrow{\sum_k}(t,T)= -\int^T_t\overrightarrow{\sigma}_k(t,u) \ du,k = n,r$
\\
20. the price at t of calls and puts on the index struck at K and expiring at T:
$V^{Index}_{\phi}(t,T,\cdot)= \phi[I(t)P_r(t,T)N(\phi h_1)-KP_n(t,T)N(\phi h_2)]$
where $\phi = 1$ for a call and $\phi = -1$ for a put.\\
21. $h_1 = (log(\frac{I(t)P_r(t,T)}{P_n(t,T)K}) + \frac{V(t,T)^2}{2}) /V(t,T).$ \\
$h_2=h_1-V(t,T)$\\
where \\
$V(t,T)^2 = \int^T_t\overrightarrow{\sum}_I(u,T)* \rho * \overrightarrow{\sum}_I(u,T) \ du$ \\
in which $\rho$ is a correlation matrix \\
defining \\
$\overrightarrow{\sum}_I(u,T) = \overrightarrow{\sum}_n(u,T)- \overrightarrow{\sum}_r(u,T)-\overrightarrow{\sigma}_I(u)$ \\
If the inflation rate between $T_1$ and $T_2$ is defined as: $R^I(T_1,T_2) = \frac{I(T_2)}{I(T_1)} - 1$, for a strike of K the option pay-off at $T_2$ is max[ $ \phi(R^I(T_1,T_2)-K),0$ ]. \\
22. The value of the option: $V^{Inflation}_{\phi}(t,T_1,T_2,\cdot) = \phi P_n(t,T_2)[K_1N(\phi j_1)-K_2N(\phi j_2)]$ \\
where: $K_1 = \frac{P_r(t,T_2)P_n(t,T_1)}{P_n(t,T_2)P_r(t,T_1)}e^{\Omega (t,T_1,T_2)}$,
$K_2 = 1 + K$ \\
23. $\Omega (t,T_1,T_2) = V(t,T_1)^2 - \int^{T_1}_t \overrightarrow{\sum}_1(u,T_1)*\rho * [\overrightarrow{\sum}_I(u,T_2)-(\overrightarrow{\sum}_n(u,T_2)-\overrightarrow{\sum}_n(u,T_1))] \ du$ \\
24. $j_1 = \frac{log(K_1/K_2) + V(t,T_1,T_2)^2/2}{V(t,T_1,T_2)}$,
$j_2 = j_1 - V(t, T_1, T_2)$
25. $V(t, T_1, T_2)^2 = \int^{T_2}_t \overrightarrow{\widetilde{\sum}}_I(u,T_1,T_2) *\rho * \overrightarrow{\widetilde{\sum}}_I(u,T_1,T_2) \ du $
and\\
$\overrightarrow{\widetilde{\sum}}_I(u,T_1,T_2) = \overrightarrow{\sum}_I(u,T_2)-H(T_1-u) \overrightarrow{\sum}_I(u,T_1)$

\subsection{Equivalent measure}

Two probability measures are said to be equivalent if they define the same null sets. The Girsanov theorem characterizes the transformation of semi-martingales under equivalent changes of measure. \cite{Encyclopedia}

In finance, the methodology of changing measures contributes to providing a bridge to connect the statistical market data to mathematical models. \\
The statistical measure, denoted by $P$, reflects the real-world dynamics of financial assets. The P-measure is a method of measuring probability based on historical data rather than based on assumptions of the existence of a risk-free rate and absence of arbitrage in the market. The P-measure has contributions in risk measurement. Moreover, the current value of a financial asset is the sum of the expected future payoffs discounted at their own rates reflecting the risk of the asset. Statistical and econometric tools are used to estimate the rates and predict the future value of assets.\\
A risk-neutral measure(Equivalent Martingale Measures), denoted by $Q$, is the measure of choice for valuation of derivative securities. \\ Q-measure is a method of measuring probability such that the current value of a financial asset is the sum of the expected future payoffs discounted at the risk-free rate(The risk free rate is the return on investment on a risk-less asset.). Q-measure has contributions in the pricing of financial derivatives under the assumption that market is free of arbitrage. 
In mathematical models, prices of traded assets are supposed to be (local) Q-martingales, and hence their dynamics(as seen under Q) differ from their actual behavior(as modeled under P).There is certainly a gap between P-probability space and Q-probability space. \\
\begin{definition}
 Let P,Q be two probability measures defined on a measurable space ($\Omega, \mathcal{F}$). We say that Q is absolutely continuous with respect to P,denoted by Q $\ll$ P, if all P-zero sets are also Q-zero sets. If Q $\ll$ P and P $\ll$ Q, we say that P and Q are equivalent, denoted by P $\sim$ Q. In other words, two equivalent measures have the same zero sets. \\
\end{definition}
Let Q $\ll$ P. BY the Radon-Nikodym theorem there exists a density $Z = dQ/dP$ so that for $f\in L^1(Q)$ we can calculate its expectation with respect to Q by 
$Z_t = E_P[Z|\mathcal{F}_t]$ \\
We call the martingale $Z = (Z_t)$ the density process of Q. The Bayes formula tells us how to calculate conditional expectations with respect to Q in terms of P. Let $0 \leq s \leq t \leq T$ and $f$ be $\mathcal{F}_t-measure$ and in $L^1(Q)$. We then have\\
$Z_sE_Q[f|\mathcal{F}_s]=E_P[Z_tf|\mathcal{F}_s]$ \\
As consequence of Baye's formula, we get that if $M$ is a Q-martingale then $ZM$ is a P-martingale and $vice \ versa$. Hence, we can turn any Q-martingale into a P-martingale by just multiplying it with the density process. It follows that the martingale property is not invariant under equivalent measure change. \\
Some important objects like stochastic integrals and quadratic variations retain invariant under equivalent measure changes, although they depend on some probability measure by their definition. This can be illustrated by a concrete example of  quadratic variation of a semi-martingale $S$. This is defined to be the limit in P-probability of the sum of the squared S-increments over a time grid, for vanishing mesh size. Since the convergence in P-probability implies convergence in Q-probability if $Q \ll P$ , the convergence in P-probability is equivalent to the convergence in Q-probability when P and Q are equivalent. This implies that quadratic variations remain the same under a change to an equivalent probability measure. \\

In addition, the changing of measures has a broad application and not only restricted to changing between Q-measure and P-measure. For example, in some models, a measure only consists of the bonds with the same maturity date. The bonds with distinct maturities are under different measures.

\section{The History of developed Models}
\subsection{Perspective of Models}
\subsubsection{Empirical Models vs Theoretical Models}

The term structure of interest rates refers to the quantified relationship between the interest rate(or the yield) and the evolution of financial instrument until the maturity. In order to describe the quantified the yield curve, as well as the relations among different financial instruments, different types mathematical models are introduced and analyzed.\cite{Lee}\\

In the past few decades, the development of interest rate models has intrigued attention from researchers in academy. Depending on the methodology of behind each model, the interest models can be classified into two categories: empirical model and theoretical model. The empirical model does not reserve the property of arbitrage-free. They involve along the price curve of desired financial instruments, which means the parameters are derived explicitly from the data set. On the contrary, the theoretical model approaches this problem from a pure mathematical perspective. Though it is derived from the observation of the market, does not have dependence on the data set. Meanwhile, it has the property of arbitrage-free and reserve its meaning along the intrinsic mathematical proofs.\\

From the perspective of empirical models, the pricing model(bond pricing model) has assumption that it is only subject to \textit{the short rate}. The model has the form of stochastic differential equation with the assumption of distribution as normal, log-normal, or non-central chi-square distribution.\cite{Brigo} The one-factor short rate models attempt to generate a process to describe movements of the yield curve by only using the short-term interest rate. A realistic method to approach the same problem would include the consideration of additional factors, such as (but not only subject to) the long term interest, the difference between short term interest rate and long term interest rate, the mean of short rate, and the volatility of short term rate. \\

From the perspective of theoretical model, the construction of no-arbitrage(arbitrage-free) is the main goal that we look for. Based upon the short rate, the discount bond prices, and instantaneous forward rates, a stochastic model is expected to have the consistence of arbitrage-free at any time point along the pricing curves. In our case, the price of nominal coupon bond should agree with the price of Inflation-indexed bond at each maturity date, such as 1 year, 2 years, 5 years and 10 years. Thus, though there is a break-even curve between the nominal bond pricing curve and the inflation-indexed bond pricing curve, the integration of the break-even curve should compensate to zero at each maturity, as the expected price of nominal bond and expected price of inflation-indexed bond equivalent to each other at each maturity.\\

\subsubsection{Multi-factor Models}
Another way to classify the type of models is by the number of factors existing in the models. The factors in the model have revealed the uncertainty to simulate the term structure. The advantage of single-factor models is to generate a perfect correlation among the prices at different spots. Meanwhile, the advantage of multi-factor models indicates the movement of term structure along time. Thus, the relationship among all the adjusted factors is expected to be computed explicitly in order to reserve the property of abitrage-free.\cite{Musiela}

A factor model has the format: 
\begin{equation}
    \label{Factor Model}
    r_t = \alpha_0 + f_{1t}b_1 + f_{2t}b_2 + ... + f_{kt}b_k + e_t 
\end{equation}
$r_t$ = interest return in period t \\
$\alpha_0$  = constant term \\
$f_{kt}$ = factor k value in period t  \\
$b_k$ = exposure \ of \ the \ financial \ instrument i to factor k \\ 
(this is also referred to as beta , sensitivity, or factor loading)\\
$e_t$ = noise for financial instrument i in period t\\
(this is the return not explained by the model)\\

Factor models has two advantages when the historical market data is applied to compute covariance and correlation. First, factor models do not require a large amount of data in order to be applied. Second, the factor model has its generality on all types financial instruments, including stocks, bonds, derivatives. However, statistical analysis is still required to ensure the accuracy of the result. In result, factor models supplies the analysis with a fit on simulating the overall covariance and correlation structure between the financial instrument and the market where it belongs to.\\
In summary of the calculating method, the parameters in the model are calculated via ordinary least squares(OLS) regression analysis. Regression analysis reveals the relationship among multi-variables by using historical data. The ideology is to assume that the relationship or trend existed in the past continues to exist in the present and future.

\subsubsection{Levy Process}
From the perspective of Geometric Brownian motion(which is a special case in Levy process), there are two types of interest rate models, which are the type of GBM-based(Geometric Brownian Motion) model and the type of rational models.  \\
The GBM-based model developed by Hughston (1998) considers the consumer price index and the nominal and real interest rate systems as the jointly driven by a multi-dimenional Brownian motion. In methodology, GBM-based models treat CPI as a foreign exchange rate. Similarly, the real interest rate system is treated as if it were the foreign interest rate system associated with the foreign currency. In the model developed by Jarrow and Yildirim(2003), this methodology has been adapted with a three-factor model(i.e.,driven by three Brownian motions). The CPI is modelled as a geometric Brownian motion, with the time-dependent volatility. The two interest rate systems are considered as extended Vasicek-type (or Hull-White) models.Similarly to Jarrow and Yildirim(2003), Dodgson and Kainth(2006) applied their short-rate approach where both nominal and inflation rates are developed by Hull-White processes. The advantage of GBM-based model is that it develops the framework of how to imply the volatility. However, the limitation of GBM-bases model has been that it does not reproduce volatility smiles. Thus, the rational model stands out for its productivity on simulating the volatility smiles.

\subsection{Pricing Model}
The most representative pricing model(no-path-track) is Black-Scholes.
\subsection{Forward Rate model}
\subsubsection{Merton}
Merton provokes a one-factor model to describe the price of discount bond with the only uncertainty as the short rate.\cite{Merton} \\
The stochastic process of Merton model is summarized as a Brownian motion process with a drift parameter. In Merton model, the assumption includes that the risk of market price is a constant.\\

$E = V_tN(d_1)-Ke^{-r \Delta T} N(d_2)$\\
where $d_1 = \frac{ln\frac{V_t}{L}+(r+\frac{\sigma^2_V}{2})\Delta T}{\sigma_V \sqrt{\Delta T}}$ \\
$d_2 = d_1 - \sigma_V \sqrt{\Delta t}$\\
$E$ = Theoretical value of a company's equity \\
$V_t$ = Value of the company's assets in period t \\
$L$ = Value of the company's debt(liability threshold) \\
$t$ = Current time period \\
$T$ = Future time period \\
$r$ = Risk-free interest rate \\
$N$ = Cumulative standard normal distribution \\
$e$ = Exponential term \\
$\sigma$ = Standard deviation of stock returns \\
The Merton Model approach uses single-point calibration and requires values for the equity, debt, and volatility ($\sigma_E$) \\
This approach solves for ($V_t,\sigma_V$) using a 2-by-2 system of nonlinear equations. We can solve $\sigma_E$ by $\sigma_V$: \\
$\sigma_E = \frac{V_t}{E}N(d_1) \sigma_V$ \\
The Merton by time series approach requires time series for equity. If the equity time series has $n$ data points, this approach calibrates a time series of n asset values $V_1, ... ,V_t$ that solve the following system of equations: \\
$E_1 = V_1N(d_1)-L_1e^{-r_1T_1}N(d_2))$ \\
$...$\\
$E_n=V_nN(d_1)-L_ne^{-r_nT_n}N(d_2)$ \\
The function directly computes the volatility of assets $\sigma_V$ from the time series $V_1,...,V_n$ as the annualized standard deviation of the log returns. This value is a single volatility value that captures the volatility of the assets during the time period spanned by this time series.  \\
After computing this values of $v_t$ and $\sigma_V$, the function computes the distance to default (DD) is computed as the number of standard deviations between the expected asset value at maturity T and debt threshold: \\
$DD = \frac{lnV_t + (\mu_V - \frac{\sigma^2_V}{2})\Delta T -ln(L)}{\sigma_V \sqrt \Delta T}$\\
The drift parameter $\mu_V$ is the expected return for the assets, which is equal to risk-free interest rate in this thesis, or any other value based on expectations for the firm. \\
The probability of default(PD) is defined as the probability of the asset value below denotes the relation with liability threshold at the end of the time horizon T. \\
$PD = 1 - N(DD)$ \\

\subsubsection{Vasicek}
The Vasicek model is a partial-empirical one-factor short rate model that describes the evolution of interest rate. It describes interest rate movement by driving only one source(which is the short rate) of market risk. The Vasicek model is used in the estimation of interest rate derivatives, as well as inflation-indexed derivatives. Though the Vasicek model can not provide the price of bonds in a closed form, it can generate an explicit solution under the specific case where the risk of market price is assumed to be constant. The assumption behind this case is called Ornstein-Uhlenbeck process, which implies a mean-reverting Gaussian interest rate curve. However, the possibility of negative interest rate retains in this model while it has been broadly used in pricing bond options.\cite{Karolyi} \\

$dr_t = a(b-r_t) dt + \sigma dW_t$ \\ 
\textbf{where:} \\ 
W = Random market risk (represented by a Wiener process)\\ 
t = Time period \\ 
a = Speed of the reversion to the mean, which is the value of b \\ 
b = The level of mean in long term, all future trajectories of r will regroup around b \\ 
$\sigma$ = Volatility at time t. The volatility is updated by time t instantaneously.  \\
W = Random market risk (represented by a Wiener process) \\
t = Time period \\
$a(b−r_t)$ = Expected change in the interest rate at time t (the drift factor) \\
$\frac{\sigma^2}{2a}$ = Variance in long term, $\sigma^2$ and $2a$ have negative correlation, because $ \sigma$ contributes toward the randomness of the system and a contributes to the speed of convergence to the mean value in long term \\

In Merton model and Vasicek model, the short rate is able to be a negative value due to the Gaussian distribution as a part in the interest rate process. Dothan renovated a log-normal distributed process, thus the short rate is strictly positive.\cite{Dothan} Unfortunately, Dothan's renovation does not provide an explicit formula for the pricing. Thus, the only way to approach the valuation and solution in Dothan's renovation is by numerical methods.\cite{Musiela}\\

\subsubsection{Cox-Ingersoll-Ross(CIR)}
CIR model is introduced in the section of Ito process, thus we skip the introduction here.
CIR model supplies a solution to the problem of negative interest rate in Merton model and Vasicek model by provoking a different one-factor short rate model where the interest rate process is under non-central chi-square distribution. \cite{cox} Due to the feature that the volatility of short rate is correlated with the $\textit{square-root process}$, the volatility is strictly positive. The closed form solution of zero-coupon bond and European call options are given by the model. More importantly, CIR model is ubiquitous in pricing methods, such as mortgage-backed security valuation model, futures and future option pricing models, swap pricing model, and yield option valuation model.\cite{chan} 
Instead of single square-root process, Longstaff advocates the CIR framework by renovating it into a $\textit{double sqaure-root process}$ for the short rate. \cite{Longstaff}\\
Brennan and Schwartz also improvised a two-factor model of interest rate. In this model, both short rate and long rate are involved and follow a log-normal distribution.\cite{Brennan}Meanwhile, Longstaff and Schwartz provoke a two-factor method with assumption of that underlying factors(dependence) as short rate and the volatility.\cite{Longstaff2} \\
Thus, the underlying factors are indeed the choice made by the researchers who construct the model. As mentioned in the summarizing work of two-factor models by Subrahmanyam, different two-factor models conclude different underlying factors, such as short rate and the spread between short rate and long rate, the short-term forward rate and inflation rate, the short rate and its mean.\cite{Subrahmanyam}
\subsection{Inversion Model}
\subsubsection{Hull-White Model}
The Hull-White model can be categorized in two cases, one-factor model and two-factor model. \\
In the case of one-factor model, the equation is: \\
$df(r(t)) = [\theta(t)-\alpha(t)f(r(t))] \ dt + \sigma(t) \ dW(t)$ \\
$\theta(t)$: a term-structure selected in order to let the model fit the initial condition. \\
$a(t)$ and $\sigma(t)$: volatility parameters that are chosen to fit the current market.To be careful, the functions $a(t)$ and $\sigma(t)$ are chosen to fit in the model. Thus, these two volatility parameters are expected to be decided before the calibration process. \\
$dW(t)$: a standard Wiener process with a zero mean and a variance equal to $dt$. \\

Essentially, Hull-White model contains other models by defining its special cases. For example, when $f(r(t)) = r(t)$, $a(t) = 0$, and $\sigma$ is a constant, the one-factor Hull-White model becomes the Ho-Lee short rate model, which is $dr(t)=\theta(t) \ dt + \sigma \ dW(t)$.(In Ho-Lee's model, a typical binomial lattice methodology is applied, which is addressed in the part of SDE's solution.) In the case of that $f(r(t))=r(t)$ and $a(t) \neq 0$, it becomes the original Hull-White model(1990). In both Ho-Lee and original Hull-White model, future interest rates of all maturities are normally distributed and analytic solutions exist for the price of bonds and derivatives. Moreover, there are other cases as well. When $f(r(t)) = \sqrt{r(t)}$, it is Pelsser model(1996). When $f(r(t))=lnr(t)$, the model is Black-Karasinski model(1991), in which future short rate is lognormally distributed and rates of all other maturities are close to lognormal distribution. In addition, in the case $f(r(t)) = ln(r(t)) $ and $\alpha = -\frac{{\sigma_t}^{'}}{\sigma_t}$, which is $dln(r(t))$ = $[\theta_t + \frac{{\sigma_t}^{'}}{\sigma_t}ln(r(t))] \ dt +\sigma_t \ dW_t$, it becomes Black-Derman-Toy model, which was initially developed by Goldman Sachs. \\

where $r$ = the instantaneous short rate at time t\\
$\theta_t$ = value of the underlying asset at option's expiration\\
$\sigma_t$ = instant short rate volatility\\
$W_t$ = a standard Brownian motion(Wiener process) under a risk-neutral probability measure.)\\
In the case of two-factor model, Hull-White model becomes:\\
$df(r(t)) =  [\theta(t) + u -a(t)f(r(t))] \ dt + \sigma_1(t) \ dW_1(t)$ \\
where u has the process below with initial value of 0:
$du = - bu \ dt + \sigma_2 dW_2(t)$ \\
In addition, approaching to the solution from pure mathematical perspective, if $a,\theta, and \sigma$ are constant, Ito lemma can be used to get:\\
$r(t) = e^{-at}r(0) + \frac{\theta}{a}(1-e^{-at})+ \sigma e^{-at}\int^t_0 e^{au} \ dW(u)$ \\
Meanwhile, $r(t)$ has distribution: \\
$r(t)~ N(e^{-at}r(0)+\frac{\theta}{a}(1-e^{-at},\frac{\sigma^2}{2a}(1-e^{-2at})))$\\
where $N(\mu,\sigma^2)$ denotes the normal distribution with mean $\mu$ and variance $\sigma^2$\\
When $\theta(t)$ is time dependent, \\
$r(t)=e^{-at}r(0)+\int^t_0 e^{a(s-t)} \theta(s) \ ds + \sigma e^{-at} \int^t_0 e^{au} \ dW(u) $\\
Meanwhile, $r(t)$ has distribution \\
$r(t) ~ N(e^{-at}r(0) + \int^t_0 e^{a(s-t)} \theta(s) \ ds, \frac{\sigma^2}{2a}(1-e^{-2at}))$
\cite{JohnHull}

\section{Korn\&Kruse Model}

\subsection{Fundamental Setups}
In the model developed by Korn and Kruse, the price index is so called Monetary Union Index of Consumer Prices(MUICP), which is a weighted average of European monetary union countries' Harmonised Index of Consumer Prices(HICP). Harmonized Index of Consumer Price(HICP) is a consumer price index which is compiled based on a methodology that has been harmonised across EU countries. The difference on mainstream between HICP of EU and CPI of US is from two primary aspects. First, HICP incorporates both urban and rural consumers. Secondly, HICP excludes owner-occupied housing, since HICP considers it as investment rather than consumption.\cite{JohnHull}
\subsection{Terms}
$r_N(t)$: nominal interest rate for an investment horizon until time t. \\
$r_R(t)$: real interest rate until time t, which is the gain in real purchasing power of an investment. \\
$\mathbb{E}[i(t)]$: expected inflation rate. \\
We have the relation: $r_N(t) = r_R(t) + E[i(t)]$
\section{HJM model}
Compared to the previous models belonging to the categories of pricing model, forward rate model, and inversion model, HJM belongs to the category of interest rate structure model.
From the perspective of mathematics, the maturity of a bond is a martingale and is equal to 1. Thus, under the assumption of no-arbitrage market, the premium factor in the pricing model of a bond does not exist. Thus, in the traditional Q measure, the price of a bond is given as:\\
$\beta(t) = \exp{\int^t_0 r_u \ du}$
\section{Jarrow-Yildirim Model}
In summary, the model developed by Jarrow-Yildirim is a path-trackable forward-rate model, with implementation of HJM model in its simulation of nominal and real zero-coupon bond prices, and CPI-U. 
\subsection{Setups}
The model developed by Jarrow and Yildirim prolongs HJM model. The purpose of Jarrow-Yildirim model is to renovate and apply HJM model to TIPS, conventional U.S. Treasury bonds, as well as related derivative securities. 

The fundamental assumption of the economy in this model is no-arbitrage , where nominal dollars correspond to the domestic currency, real dollars correspond to the foreign currency, and the inflation index corresponds to the spot exchange rate. \\
The first step of calculation is to strip the nominal and real zero-coupon bond prices by applying the prices of the coupon-bearing conventional U.S. Treasury bonds and TIPS, respectively. The best fitting piece-wise constant forward rate curve is obtained by using a nonlinear least square algorithm.\\
The second step is letting the price curves of time-series evolutions of the CPI-U, the real zero-coupon bond, and nominal zero-coupon bond to adjust the three-factor HJM model.\\
The third step is to utilize the estimated parameters to test the validity of HJM model through its hedging performance in the secondary market for TIPs. In return, the hedging analysis assures the validity of the three-factor extended Vasicek model.\\
The fourth step is to demonstrate the effect of this model by pricing a call option based on the CPI-U inflation index. Though this call option is not traded, such an option is constructed in a closed form by using standard hedging procedures in this complete market model.\\
\subsection{Hypotheses}
1.The most important ideology of hypothetical cross-currency economy introduced by Jarrow-Yildirim model is to treat nominal dollars as domestic currency, to treat real dollars as the foreign currency, and to treat the inflation index as the spot exchange rate. \\
2.The equation $B_n(0) = \sum^T_{t=1}CP_n(0,t) + FP_n(0,T)$ below is a no-arbitrage restriction that holds under the standard frictionless and competitive market hypotheses. In particular, it is assumed that there are no transaction costs, no restrictions on trades, and no differential taxes on coupons vs. capital gains income.(There is some evidences: the differential state taxes on corporate vs. government bonds may be important for the determination fo corporate bond yields(see Elton, Gruber, Agrawal, and Mann(2001))).\\
\subsection{Terms}
The real price of a derivative corresponds to the foreign currency price, the nominal price of a derivative corresponds to the domestic prices, and the CPI-U index corresponds to the exchange rate. \\
r: real. \\
n: nominal. \\
$r_r$: real rate. \\
$r_n$: nominal rate.\\
$\tau$: stopping time.\\
$P_n(t,T)$: time t price of a nominal zero-coupon bond maturing at time T in dollars. \\
$I(t)$: time t CPI-U inflation index, i.e,. dollars per CPI-U unit(lagged two months). \\
$B(t)$: Numeraire at time t.(Numeraire can be considered as an operator to indicate a quantitative relationship.)
$P_r(t,T)$: time t price of a real zero-coupon bond maturing at time T in CPI-U units. \\
$f_k(t,T)$: time t forward rates for date T where $k \in \{r,n \}$  \\i.e., \\
1.$P_k(t,T) = exp [-\int^T_t f_k(t,u) \ du]$ \\
$r_k(t) = f_k(t,t): The time t spot rate where k \in \{ r,n \} $\\
$B_k(t) = exp \{ \int^t_0r_k(v) \ dv \}$: Time t money market account value for $ k \in \{ r,n \} $ \\
(2)$B_n(0)$: Time 0 price of a conventional coupon-bearing bond in dollars where the coupon payment is C dollars per period, the maturity is time T, and the face value is F dollars  \\
$B_n(0) = \sum^T_{t=1}CP_n(0,t) + FP_n(0,T)$. \\
(3)$B_{TIPS}(0) = \{ \sum^T_{t=1}CI(0)P_r(0,t) + FI(0)P_r(0,T) \} /I(t_0)$. \\
(4)$P_{TIPS}(t,T) = I(t)P_r(t,T)$. \\
We consider a continuous trading economy with trading interval $[0, \tau]$. The uncertainty in the economy is characterized by a probability space $(\Omega, \mathbb{F},\mathbb{P})$ where $\Omega$ is a state space, F is the set of possible events(a $\sigma$ on $\Omega$), and $P$ is the statistical probability measure on ($\Omega$, $\mathbb{F}$). Furthermore, let $\{ F_t: t \in [0,T] \}$ be the standard filtration generated by the three Brownian motions $(W_n(t), W_r(t),W_I(t))$: $t \in [0,T])$. These Brownian motions are initialized at zero with correlations given by $dW_n(t)dW_r(t)=\rho_{nr}dt,dW_n(t)dW_I(t)=\rho_{nI}dt$, and $dW_r(t)dW_I(t)=\rho_{rI}dt$. Hence it is a three-factor model. \\
(5)Based on the initial forward rate curve $f_n(0,T)$, the nominal T-maturity forward rate evolves as: 
$df_n(t,T) = \alpha_n(t,T)dt + \sigma_n(t,T)dW_n(t),$
where $\alpha_n(v,T)$ is random and $\sigma_n(v,T) $ is a deterministic function of time subject to some technical smoothness and boundedness conditions. The deterministic volatility in expression (5) implies that the nominal term structure of interest rates generates a Gaussian economy. \\
(6) Similarly, by given the initial forward rate curve $f_r(0,T)$, the real T-maturity forward rate evolves:\\
$df_r(t,T) = \alpha_r(t,T)dt + \sigma_r(t,T)dW_r(t)$
where $\alpha_r(t,T)$ and $\sigma_r(t,T)$ satisfy the same conditions as in expression(5).\\
(7)Most importantly, the inflation index's evolution is:\\
$\frac{dI(t)}{I(t)} = \mu_I(t)dt + \sigma_I(t)dW_I(t)$,\\
where $\mu_I(t)$ is random and $\sigma_I(t)$ is a deterministic function(based on data-set rather than a specific variable) of time subject to some technical smoothness and boundedness conditions. Since the deterministic volatility in expression (7) indicates that the inflation index follows a Geometric Brownian motion, so that the logarithm of the inflation index process is normally distributed. \\
These evolutions are arbitrage-free and the market is complete(Amin-Jarrow 2011) if there exists a unique equivalent probability measure Q such that \\
(8)$\frac{P_n(t,T)}{B_n(t)}$, $\frac{I(t)P_r(t,T)}{B_n(t)}$ and $\frac{I(t)B_r(t)}{B_n(t)}$ are Q-martingales. \\
By  Girsanov's theorem,given  that $(W_n(t),W_r(t),W_I(t): t \in [0,T])$ is a P-Brownian motion and that Q is a probability measure equivalent to P, then there exist market prices of risk $(\lambda_n(t), \lambda_r(t),\lambda_I(t): t \in [0,T])$ such that \\ (9)$\widetilde{W}_k(t) = W_k(t) - \int^t_0 \lambda_k(s) \ ds$ for  $k \in \{ n, r, I \}$ are Q-Brownian motions. \\
The stochastic processes $(\lambda_n(t), \lambda_r(t),\lambda_I(t): t \in [0,T])$ are the risk premiums for the three risk factors in the economy. \\

Moreover, the reason for (9) is explained as the following. suppose $W_k(t)$ is an $\mathbb{F_t}$-adapted process that is right continuous with left limit of which path are increasing, and suppose $W_k(t)$ is integrable for each $t$. Since $t \geq s$ and $W_k(t) \geq W_k(s)$, then $\mathbb{E}[W_k(t)|\mathbb{F_s}] \geq \mathbb{E}[W_k(s)|\mathbb{F_s}] = \mathbb{W_k(s)}]$, thus $W_k(t)$ is a submartingale. By the Doob-Meyer decomposition, there exists a predictable increasing process $\widetilde{W}_k(t)$ such that  $W_k(t)= martingale + \widetilde{W}_k(t)$. Thus, it is equation (9),$\widetilde{W}_k(t) = W_k(t) - \int^t_0 \lambda_k(s) \ ds$.\cite{RichardBass}

\subsection{Equations}

\begin{equation}
    \label{Jarrow & Yildilrim}
\noindent    dr_n(t) = [\Theta_n(t) -a_nr_n(t)]dt + \sigma_ndW_n(t) \\
dr_r(t) = [\Theta_r(t) - a_r r_r(t)-\sigma_r\sigma_I \rho_{rI}]dt +\sigma_r dW_r(t) \\
df_r(t,T) = \sigma_r(t,T)][\int^T_t\sigma_r(t,s)ds-\rho_{rI}\sigma_I(t)] \ dt + \sigma_r(t,T) \ d\widetilde{W}_r(t), \\
    \frac{dI(t)}{I(t)} = [r_n(t)-r_t(t)] \ dt + \sigma_I(t) \ d\widetilde{W}_I(t), \\
    \frac{dP_n(t,T)}{P_n(t,T)} =r_n(t) \ dt - \int^T_t\sigma_n(t,s)d\widetilde{W}_n(t),
    \\
    \frac{dP_{TIPS}(t,T)}{P_{TIPS}(t,T)} = r_n(t) \ dt +\sigma_I(t) \ d\widetilde{W}_I(t)- \int^T_t\sigma_r(t,s) \ ds \ d\widetilde{W}_r(t), \\
    \frac{dP_r(t,T)}{P_r(t,T)} = [r_r(t) - \rho_{rI}\sigma_I(t)\int^T_t\sigma_r(t,s) \ ds] \ dt - \int^T_t \sigma_r(t,s) \ ds \ d\widetilde{W}_r(t).
\end{equation}

Moreover, we obtain the following relation under Q-measure: \\

    \noindent 
    $r_n(t) = r_n(s)e^{-a_n(t-s)}+\int^t_s e^{a_n(u-t)}\Theta_n(u)du + \int^t_s e^{a_n(u-t)\sigma_n dW_n(u)}$ \\
    $r_r(t) = r_r(s)e^{-a_r(t-s)}+\int^t_s e^{a_r(u-t)}(\Theta_r(u)- \rho_{rI}\sigma_r \sigma_I)du + \int^t_s e^{a_r(u-t)}\sigma_r dW_r(u)$ \\
    $I(t) = I(t)e^{\int^T_t(r_n(s)-r_r(s))ds -\frac{1}{2}\sigma^2(T-t)+\sigma_I(W_I(T)-W_I(t))}$ \\

Furthermore, there is a numeraire to indicate the relation between initial CPI $I(t)$ with the current CPI $I(t)$ at time t.\\

The model of Jarrow and Yildilrim provides a method of evolution of the real and nominal forward rates and nominal zero-coupon bond prices. The model indicates that both real and nominal forward rates are normally distributed, and the inflation index follows a geometric Brownian motion.\\
In addition, there are two propositions given by Jarrow and Yildirim.\\
Proposition 1: The arbitrage-free term structures $\frac{P_n(t,T)}{B_n(t)}$, $\frac{I(t)P_r(t,T)}{B_n(t)}$ and $\frac{I(t)B_r(t)}{B_n(t)}$ are Q-martingales, if and only if the following condition holds: \\
$\alpha_n(t,T) = \sigma_n(t,T)(\int^T_t \sigma_n(t,s)ds - \lambda_n(t))$\\
$\alpha_r(t,T) = \sigma_r(t,T)(\int^T_t \sigma_r(t,s)ds - \sigma_I(t) \rho_{rI} - \lambda_r(t))$\\
$\mu_I(t) = r_n(t) - r_r(t) - \sigma_I(t) \lambda_I(t)$ \\
$\alpha_n$ and $\alpha_r$ are two arbitrage-free forward rate drift restrictions on nominal rate and real rate.(The difference is that $\alpha_n$ is from the original HJM model's frame work and $\alpha_r$ is analogous.) The last equation $\mu_I(t)$ indicates the relation among nominal interest rate, real interest rate, and the expected inflation rate. \\
Proposition 2: \\
$df_n(t,T) = \sigma_n(t,T)\int^T_t \sigma_n(t,s)ds + \sigma_n(t,T)d \widetilde{W}_n(t)$, this price process holds under the martingale measure, so called "the term structure evolution under the martingale measure.

\section{Rational Kernel}
\subsection{Terms}
\noindent $(\pi_t)_{0 \leq t}$: pricing kernel \\
$P^N_{tT}=\frac{1}{\pi^N_t}\mathbb{E}^P_t[\pi^N_T]$: nominal zero-coupon bond price system \\
$C_t=\frac{\pi^R_t}{\pi^N_t}$: CPI (works as exchange rate) \\
$d C_t= C_t(r^N-r^R) dt + C_t \sigma_C dW^{C_t}$: Inflation index under the nominal risk-neutral measure\\
(ps. Jarrow Yildrim Model(2003)):\\
$dr^N_t= [\Theta_N(t)-a_N r^N_t] \ dt + \sigma_N dW^N_t$\\
$dr^R_t = [\Theta_R(t) - \rho_R C^{\sigma} C^{\sigma_R} - a_R r^R_t]  dt + \sigma_R dW^{R_t}$ \\
$dC_t = C_t(r^N_t - r^R_t) dt + C_t \sigma_C dW^{C_t}$ \\
where$r^N_t$ denotes the nominal rate curve,$r^R_t$ denotes the real rate curve, $C_t$ denotes the CPI curve, $W^N_t$,$W^R_t$,and $W^C_t$ are dependent Brownian motions, and $\theta_N(t)$ and
$\Theta_R(t)$ are functions chosen to fit the term-structure of interest rates.)\\
$h^R_t=\frac{\pi^R_t}{M_t}$: real pricing kernel $(h^R_t)$ under $M_t$ \\
$h^N_t = s_t h^R_t$: nominal pricing kernel \\
($s_t= \frac{1}{\pi^R}\exp(- \int^T_0 r^N_s ds)\frac{dQ^N}{dP}|_{\mathbb{F}_t}$) \\
$h^R_t=R(t)[1+b^R(t)(A^R_t - 1)]$: rational pricing kernel\\
(R(t) is deterministic function) \\
$Q^N$: nominal risk-neutral measure \\
$Q^R$: real-risk neutral measure \\
Benchmark model: Geometric Brownian Motion specification of Black-Scholes \\
$(V^{\chi}_t)_{0<t<T}: price process$\\
$\chi = a_1 + a_2 C_t$: payoff function \\
($P^N_{tT} = \frac{b_2(T)A^S_t + b_3(T)\mathbb{E}^\mathbb{M}_t[A^R_T A^S_T]}{b_2(t)A^S_t+ b_3(t)A^R_t A^S_t}$) \\
$P^{IL}_{tT}=\frac{b_0(T) + b_1(T) A^R_t}{b_2(t) A^S_t + b_3(t)A^R_t A^S_t}$: price of inflation-indexed bond - zero coupon. \\
Change of Measure: $\mathbb{E}^M_t[h_T X] = \mathbb{E}^Q_t[Z_T X]$\\
$(r_t)$: short rate \\
$(h_t)$decomposition: $h_t = \int^t_0 - h_{s-r_{s-}} ds + \int^t_0 \frac{h_{s-}}{\varepsilon_{s-}}d\varepsilon_s$ \\
$(r^R_t)_{0 \leq t}$: real short rate \\
$P^R_t = exp(-\int^t_0 r^R_s ds)$: discount factor\\
$I^R_t = \int^t_0 \frac{b^R(S)}{1+b^R(s)(A^R_{s-}-1)}dA^R_s$\\
Change of Measure (like Girsanov Theorem): $\varepsilon^R_t = \frac{dQ^R}{dM}|\mathbb{F}_t$
\cite{Henrik}
\subsection{Rational Pricing Kernel System}

Rational pricing kernel system is one of the methods to calculate and predict the prices of zero coupon swap and Year-on-Year swap, which serve to hedge instruments against future risk or loss on asset. \\

 (Rational Pricing Kernel System): Let $\mathcal{M}$ be a \textit{measure equivalent} to $\mathcal{P}$ induced by a \textit{Radon-Nikodym} process $({M_t})_{0\leq t}$. Let $(A^R_t)_{0 \leq t}$ and $(A^S_t)_{0 \leq t}$ be \textit{unit-initialised} and \textit{positive martingales} under $\mathcal{M}$. Let $(A^R_t)$, $(A^S_t)$ and  $(A^R_tA^S_t)$ be $\mathcal{M}$-integrable for all $t \geq 0$. Let the real pricing kernel $(h^R_t)_{0 \leq t}$ br defined by

\centerline{$h^R_t=R(t)[1+b^R(t)(A^R_t-1)]$}
\noindent 
where $R(t) \in C^1$ is a $\textit{unit-initialised}$ and strictly positive deterministic function, and where $b^R(t) \in C^1$ is a deterministic function that satisfies $0 < b^R(t) < 1$.

Furthermore, let \\

\centerline{$s_t=S(t)A^S_t$}
where $S(t) \in C^1$is unit-initialised and strictly positive deterministic function, and set\\ 
\centerline{$h^N_t=s_t h^R_t$}.\\
We call $(h^R_t,s_t,h^N_,M_t)_{0 \leq t}$ thus specified a rational pricing kernel system(RPKS).

\section{My model}
\subsection{Motivation}

\begin{theorem}
A market model is arbitrage-free if and only if it has a risk-neutral probability measure. This is the fundamental theorem of asset pricing.
\end{theorem}

\subsection{Theoretical Model}
In our model, CPI is defined as a function $I(t)$ as the only dependence is time $t$.\\
Following next, the inflation rate over the time period $[t,T]$ is defined as the percentage change of the index, $\widetilde{i}(t,T)=\frac{I(T)}{I(t)}-1$ \\
Furthermore, the annualized inflation rate is defined as:\\
$i(t,T)=\frac{1}{T-t}(\frac{I(T)}{I(t)}-1)$

The zero-coupon inflation-indexed swap(ZCIIS) defines a swap contract between two parties with a single exchange of payments. Form the contract starting at time $t$ to the expiration time at time $T$, one party pays the fixed amount:\\
$(1+K(t,T))^{T-t} -1$ \\
where $K(t,T)$ is the quote for the contract. \\
The other party pays the floating rate, which is the inflation rate over the contract time:\\
$I(t,T)$ \\
The underlying asset of ZCIIS is the zero-coupon bond of which price varies to the current inflation rate. \\
The zero-coupon bond has the price:\\
$X_{ZC}(t,T) = \mathbb{E^Q}[e^{-\int^T_tr_s ds} \frac{I(T)}{I(t)}|\mathcal{F}_t]=X(t,T)(1+K(t,T))^{T-t}$ \\
where $r_s$ denotes $r(s)$ as the function of r and parameter of s.\\
Thus, the price of bond $X(t,T)$ is also considered as a discounted factor. \\
Meanwhile, for the zero-coupon bond with the same maturity date T however an earlier issuance date $T_0<t$, the price is defined:
$X_{ZC}(t,T_0,T)=\mathbb{E}^Q[e^{-\int^T_t r_s ds }\frac{I(T)}{I(T_0)}] = \frac{I(T)}{I(T_0)}X_{ZC}(t,T)$\\
Since Zero-Coupon bond's underlying is inflation rate, of which underlying is CPI, thus the CPI is defined as the ratio of change on inflation rate. CPI is defined by:\\
$\frac{I(T)}{I(T_0)}=e^{\int^T_{T_0} i(s)ds}$,\\
where $i(t)$ denotes the instantaneous inflation rate.

In most of cases, the theoretical has the structure based on $e^{-\int^T_t r(s) ds}$, since this structure offers a solid and sound interpretation not only on the phenomenon that the price growth follows an exponential growth speed, but also on the computation to get a converted rate by division of two exponential numbers.
Sometimes, the structure based on$\int^T_t e^{r(s)}ds$ also offers an approachable structure. However, the interval of integration requires to be unified before summing up two different terms of integration, which has the potential possibility to cause extra computation.\\
\subsubsection{Numeraire}
\noindent The first step: \\
Defining a numeraire B as:\\
$dB = r*B \ dt$ \\ where $r$ could be a stochastic process. \\
A classic example would be: $B(T) = e^{-\int^T_0 r(s) ds}$\\
The second step:  \\
Finding the transaction between statistical measure $P$ and the risk neutral measure $Q$ on the selected instrument $B$ : $\mathbb{P}^B \sim \mathbb{Q}^B$ \\
The third step:\\
To build a correlation under the risk neutral measure among the specific instrument $V$ and the any given set of derivatives $P$:
$\mathbb{E}^Q(\frac{P(T)}{V(T)}) = \frac{P(t)}{V(t)}$ \\
where $T$ denotes the maturity time and $t$ denote the current time. \\
Since we have constructed the risk-neutral measure for $B$, so the value of $V(t)$ and $V(T)$ are known, as well as the current value $P(t)$. \\
Thus, we are seeking the relation between $\frac{P(t)}{V(t)}$ and $\frac{P(T)}{V(T)}$\\
(An example could be: $\frac{P(0)}{V(0)}$ = $e^{-\int r(t) dt} \mathbb{E}[\frac{P(T)}{V(T)}]$ where the correlation curve is $e^{-\int r(t) dt}$)\\
Thus, the goal is to use different techniques to construct an evolving curve of r, to generate the value of $P(T)$. \\
The third step can be interpreted from the perspective of Black-Scholes as: we have known the price of the chosen instrument $V(T)$ and the price of the asset $P(T)$ at time $T$, we also have known the price of the chosen instrument $V(t)$ at time $t$, we are evaluating the price of $P(t)$. The process of valuing $X(t)$ can be considered as seeking for the estimated price of the option $P(t)$ at time $t$.

\subsubsection{term}
\noindent $I(t)$: CPI at time t \\
$i(t)$:Inflation rate at time t \\
$r_r$:real rate\\
$r_n$: nominal rate\\
$P_{ZC}(t,T)$:The price of nominal bond starting at time t and with maturity time T\\
$P_{IIB}$:The price of inflation-indexed bond \\
$BR$:The break-even (difference on price between the nominal bond and the inflation-indexed bond) \\
$SWAP_{fixed}$: Fixed leg in an inflation swap \\
$SWAP_{floating}$: Floating leg in an inflation swap \\
$Q$: Risk Neutral measure \\
$P$: Statistical measure \\

\subsection{Semi Q-measure}
The beauty of Semi Q-measure really comes along two ways: the definition and the application. While there are prerequisites for the setups, it is still Q-measure from the definition of category. Meanwhile, the statistical data from market can be applied to the model in order to predict future movements.\\
The main obvious difference compared to the previous case, where we define CPI as $\frac{I(T)}{I(T_0)}=e^{\int^T_{T_0} i(s)ds}$, now we define CPI as $\frac{I(T)}{I(t)}= e^{\int^T_t(r_n(s)-r_r(s))ds - \frac{1}{2}\sigma^2_I(T-t) + \sigma_I(W_I(T)-W_I(t))}$,
Thus, we have the numeraire defined as: \\
$dB^k(t)=r^k(t)B^k(t)dt$\\
$B^k(t)=e^{\int^t_0 r_k(s)ds}$ where $k=n,r$ \\
More specifically, the numeraire for the real rate $B^r(t)$ at time t is:\\
$B^r(t) = \frac{I(0)}{I(t)}e^{\int^t_0r_n(s)ds - \frac{1}{2}\sigma^2_It + \sigma_IW_I(t)}$ \\
$d(B^rI)(t)= I(t)dB^r(t) + B^r(t)dI(t)= I(t)(r_r(t)B^r(t)dt) + B^r(t)I(t)((r_n(t)-r_r(t))dt + \sigma_I dW_I(t)) = r_n(t)(B^r I)(t)dt +\sigma_I(B^r I)(t)dW_I(t)$

\subsubsection{Jarrow-Yidilrim Framework}
The prerequisite condition of Jarrow-Yildirim Model is:\\
The arbitrage-free term structures $\frac{P_n(t,T)}{B_n(t)}$, $\frac{I(t)P_r(t,T)}{B_n(t)}$ and $\frac{I(t)B_r(t)}{B_n(t)}$ are Q-martingales, if and only if the following condition holds: \\
$\alpha_n(t,T) = \sigma_n(t,T)(\int^T_t \sigma_n(t,s)ds - \lambda_n(t))$\\
$\alpha_r(t,T) = \sigma_r(t,T)(\int^T_t \sigma_r(t,s)ds - \sigma_I(t) \rho_{rI} - \lambda_r(t))$\\
$\mu_I(t) = r_n(t) - r_r(t) - \sigma_I(t) \lambda_I(t)$ \\
$\alpha_n$ and $\alpha_r$ are two arbitrage-free forward rate drift restrictions on nominal rate and real rate.(The difference is that $\alpha_n$ is from the original HJM model's frame work and $\alpha_r$ is analogous.) The last equation $\mu_I(t)$ indicates the relation among nominal interest rate, real interest rate, and the expected inflation rate. \\

Again, The model of Jarrow-Yildirim indicates that both real and nominal forward rates are normally distributed, and the inflation index follows a geometric Brownian motion.\\

Following the framework created by Jarrow-Yidilrim, we have:\\
\begin{definition}
The price of a Zero-Coupon bond under Q-measure is defined as:\\
$P(t,T) = \mathbb{E}^Q[e^{-\int^T_tr(s)ds}P(T,T)|\mathcal{F}^W_t]=\mathbb{E}^Q[D(t,T)|\mathcal{F}^W_t]$
\end{definition}
Note that the payoff of a Zero-Coupon bond at maturity T under Q-measure is $P(T,T)=1$. Meanwhile, the price of a Zero-Coupon bond at current time t under Q-measure is $P(t,T)$, which is correlated with the price at maturity by a discounted factor $e^{-\int^T_tr(s)ds}$ as  $P(t,T) = \mathbb{E}^Q[e^{-\int^T_tr(s)ds}P(T,T)|\mathcal{F}^W_t]$.
In addition, price of a Zero-Coupon bond under P-measure at current time t with maturity T is defined as:\\ $D(t,T)$\\
Thus, when the computation $P(t,T) =\mathbb{E}^Q[D(t,T)|\mathcal{F}^W_t]$ is under operation, the Girsanov's theorem will be involved, since we are transforming the price $D(t,T)$ under P-measure to $P(t,T)$ under Q-measure.
\begin{definition}
 The short rate instantaneous rate at time t is the process: \\
 $r(t) = f(t,t)$
\end{definition}
with properties:\\
1) the short rate dynamics: \\
$dr(t) = a(t)dt + b(t)dW(t)$\\
2) forward rate dynamics:\\
$df(t,T)=a(t,T)dt + \sigma(t,T)dW(t)$ \\
3)Zero-Coupon bond price dynamics:\\
$dP(t,T) = P(t,T)m(t,T)dt + P(t,T)v(t,T)dW(t)$\\

\subsubsection{Renovation on JY model}
The original verison of JY model is given as:

\begin{equation}
    \label{Jarrow & Yildilrim}
\noindent    dr_n(t) = [\Theta_n(t) -a_nr_n(t)]dt + \sigma_ndW_n(t) \\
dr_r(t) = [\Theta_r(t) - a_r r_r(t)-\sigma_r\sigma_I \rho_{rI}]dt +\sigma_r dW_r(t) \\
df_r(t,T) = \sigma_r(t,T)][\int^T_t\sigma_r(t,s)ds-\rho_{rI}\sigma_I(t)] \ dt + \sigma_r(t,T) \ d\widetilde{W}_r(t), \\
    \frac{dI(t)}{I(t)} = [r_n(t)-r_t(t)] \ dt + \sigma_I(t) \ d\widetilde{W}_I(t), \\
    \frac{dP_n(t,T)}{P_n(t,T)} =r_n(t) \ dt - \int^T_t\sigma_n(t,s)d\widetilde{W}_n(t),
    \\
    \frac{dP_{TIPS}(t,T)}{P_{TIPS}(t,T)} = r_n(t) \ dt +\sigma_I(t) \ d\widetilde{W}_I(t)- \int^T_t\sigma_r(t,s) \ ds \ d\widetilde{W}_r(t), \\
    \frac{dP_r(t,T)}{P_r(t,T)} = [r_r(t) - \rho_{rI}\sigma_I(t)\int^T_t\sigma_r(t,s) \ ds] \ dt - \int^T_t \sigma_r(t,s) \ ds \ d\widetilde{W}_r(t).
\end{equation}

Moreover, we obtain the following relation under Q-measure: \\
\begin{equation}
      \label{Jarrow& Yildilrim Inference}
    r_n(t) = r_n(s)e^{-a_n(t-s)}+\int^t_s e^{a_n(u-t)}\Theta_n(u)du + \int^t_s e^{a_n(u-t)\sigma_n dW_n(u)} \\
    \\
    r_r(t) = r_r(s)e^{-a_r(t-s)}+\int^t_s e^{a_r(u-t)}(\Theta_r(u)- \rho_{rI}\sigma_r \sigma_I)du + \int^t_s e^{a_r(u-t)}\sigma_r dW_r(u) \\
    I(T) = I(t)e^{\int^T_t(r_n(s)-r_r(s))ds -\frac{1}{2}\sigma^2(T-t)+\sigma_I(W_I(T)-W_I(t))} \\
\end{equation}

In order to renovate the model, we consider the CPI as: \\
$I(t,T) = I(t)\frac{B_r(t,T)}{B_n(t,T)}=I(t) e^{\int^T_t f_n(t,s)-f_r(t,s)ds}$ \\
We define the forward inflation rate as: $f_i(t,s)$ = $f_n(t,s) - f_r(t,s)$. It can be considered as the break-even(implied curve), the difference between nominal rate curve and real rate curve. \\
The modification further becomes:\\
CPI's evolution: $\frac{dI(t)}{I(t)}= r_i(t)dt + \sigma_I(t)dW^I_t$\\
TIPs(Zero-Coupon bond)' evolution: $\frac{dP_{TIPS}(t,T)}{P_{TIPS}(t,T)} = [r_i(t)+ \mu_{TPIs}(t,T)] dt +\sigma_{TIPS}(t,T)dW^i_t$ \\
Nominal bond's evolution:
$\frac{dP_n(t,T)}{P_n(t,T)}=[r_n(t)+\mu_n(t,T)]dt+\sigma_{P_n}(t,T) dW^n_t$ \\

Thus, the value of the drift term in TIPs' dynamic becomes:\\
$\mu_{TIPs}$= $\sigma_{TIPS}^2(t,T) - \rho_{iI}\sigma_I(t)\sigma_{TIPS}(t,T) + [\rho_{nI}\sigma_I(t)-\rho_{ni}\sigma_{TIPS}(t,T)]\sigma_{P_n}(t,T)$\\
The value of the drift term in Nominal bond's dynamic becomes:\\
$\mu_{n}= \sigma_n^2(t,T) - \rho_{iI}\sigma_I(t)\sigma_n(t,T)+[\rho_{nI}\sigma_{I}(t)-\rho_{ni}\sigma_n(t,T)]\sigma_{TIPs}(t,T)$ \\
(p.s. $\rho $ indicates the one of the correlation factors in the correlation matrices among Brownian motion $W_n$ $W_r$ and $W_i$, where$(W_t^n,W_t^i,W_t^I)$ is a Brownian motion under $\mathbb{P}_n$ with correlation matrix\\

$$B=
\begin{vmatrix}
1 & \rho_{ni} & \rho_{nI}\\
\rho_{ni} & 1 & \rho_{iI}\\
\rho_{nI} & \rho_{iI} & 1\\
\end{vmatrix}
$$

\subsubsection{Monte Carlo Simulation on Semi-Q-measure Model}
The market data is imported from "Fred Economic Data". The imported TIPS data was abbreviated as"DFII10", the imported break-even inflation is abbreviated as "T10YIE", as well as the 10 year Zero-Coupon bond(which is the nominal bond in our case).
Thus, the following python code has been run:
\includegraphics[width=1.0\linewidth,height=1.0\textheight,keepaspectratio]{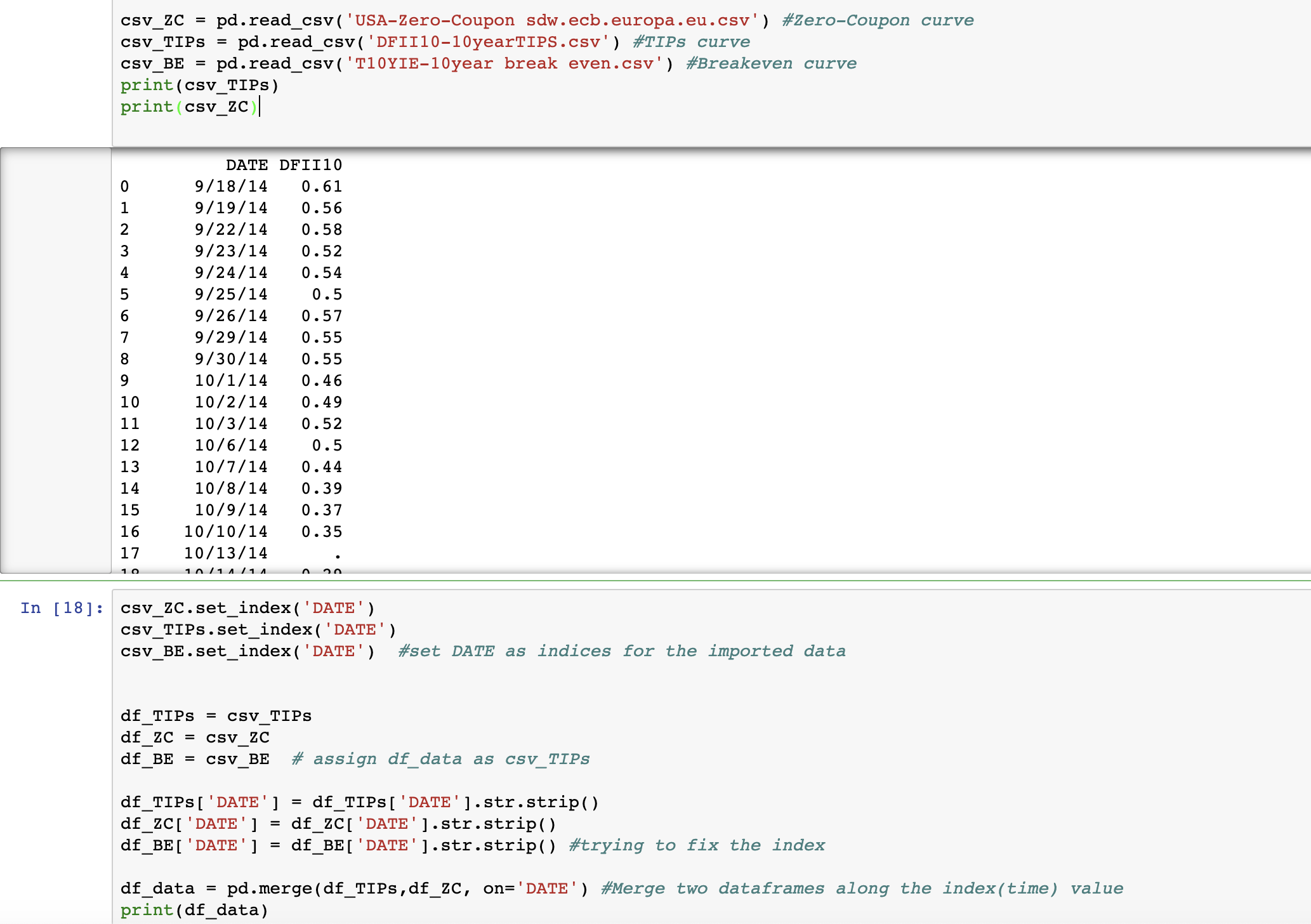}\\
\includegraphics[width=1.0\linewidth,height=1.0\textheight,keepaspectratio]{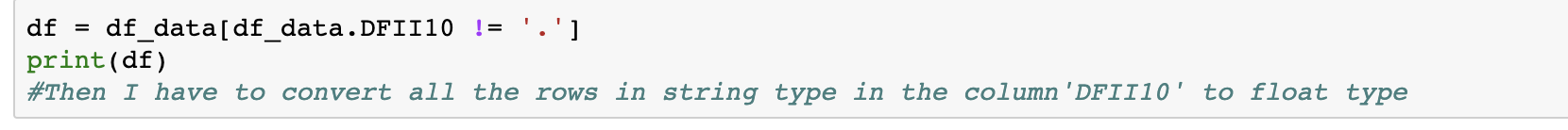}
\includegraphics[width=1.0\linewidth,height=1.0\textheight,keepaspectratio]{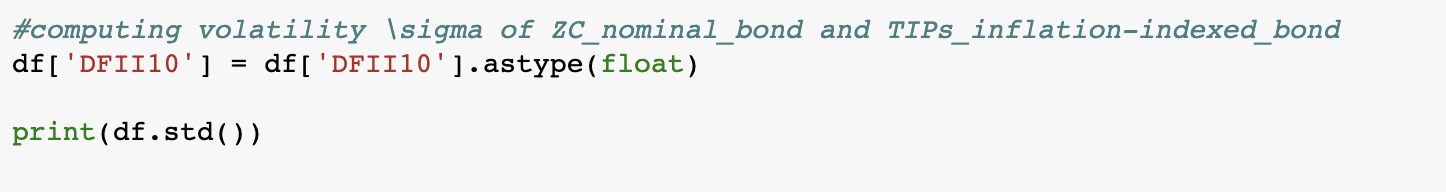}
\includegraphics[width=1.0\linewidth,height=1.0\textheight,keepaspectratio]{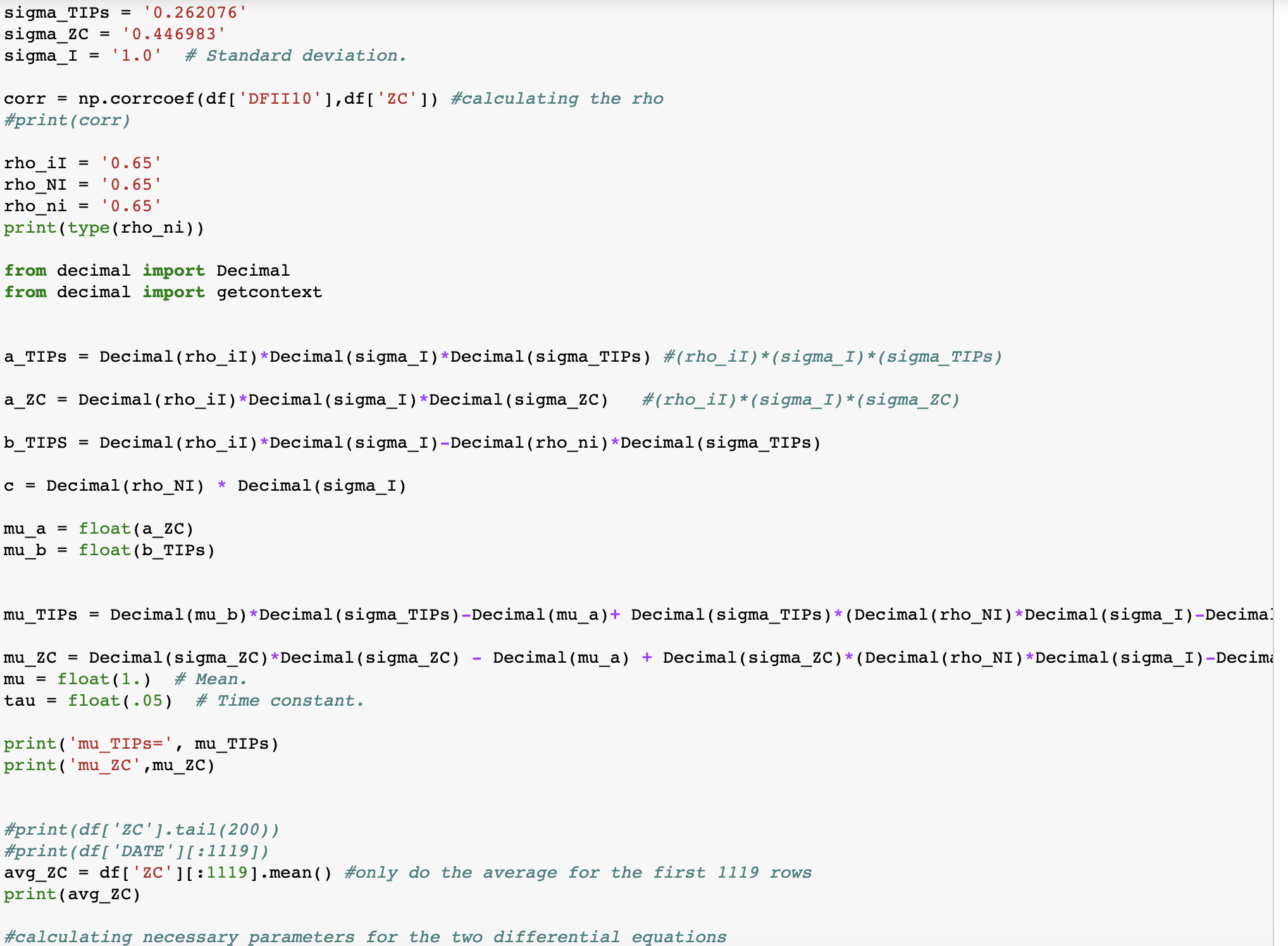}
\includegraphics[width=1.0\linewidth,height=1.0\textheight,keepaspectratio]{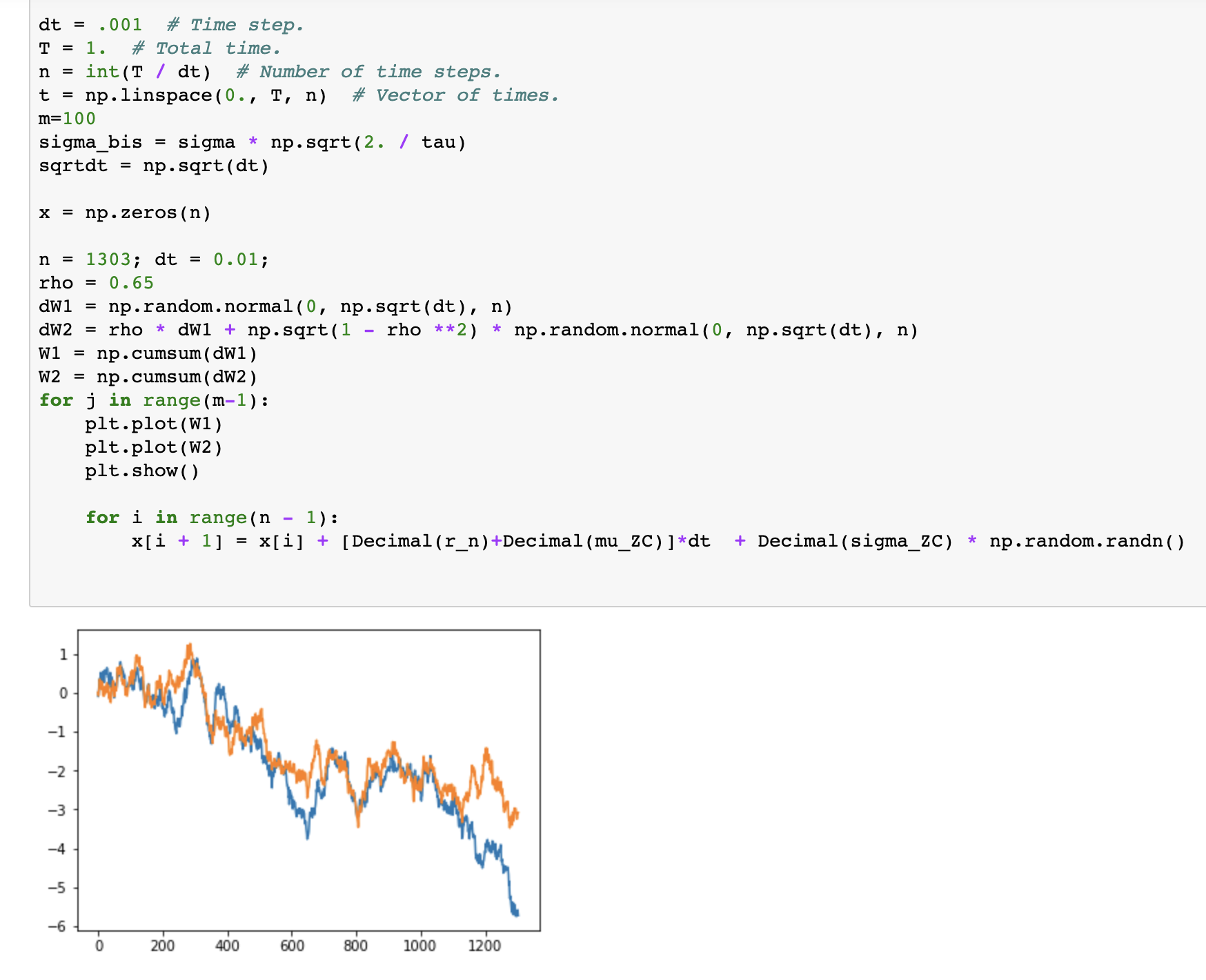}
\includegraphics[width=1.0\linewidth,height=1.0\textheight,keepaspectratio]{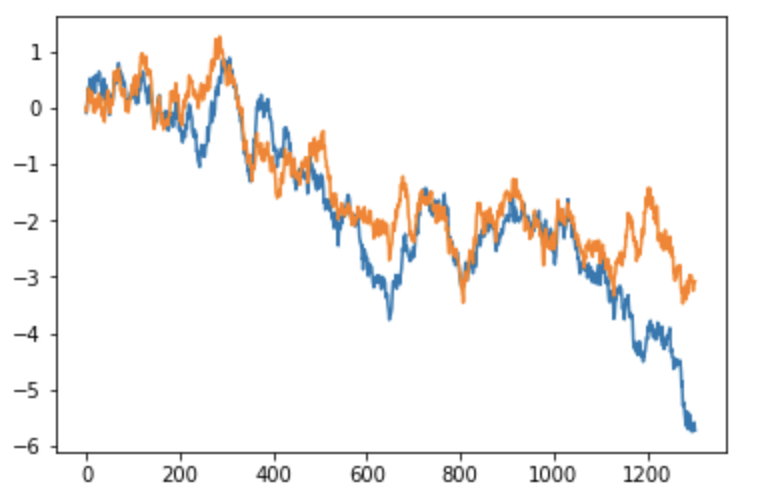}
\includegraphics[width=1.0\linewidth,height=1.0\textheight,keepaspectratio]{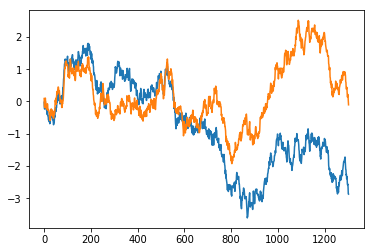}
\includegraphics[width=1.0\linewidth,height=1.0\textheight,keepaspectratio]{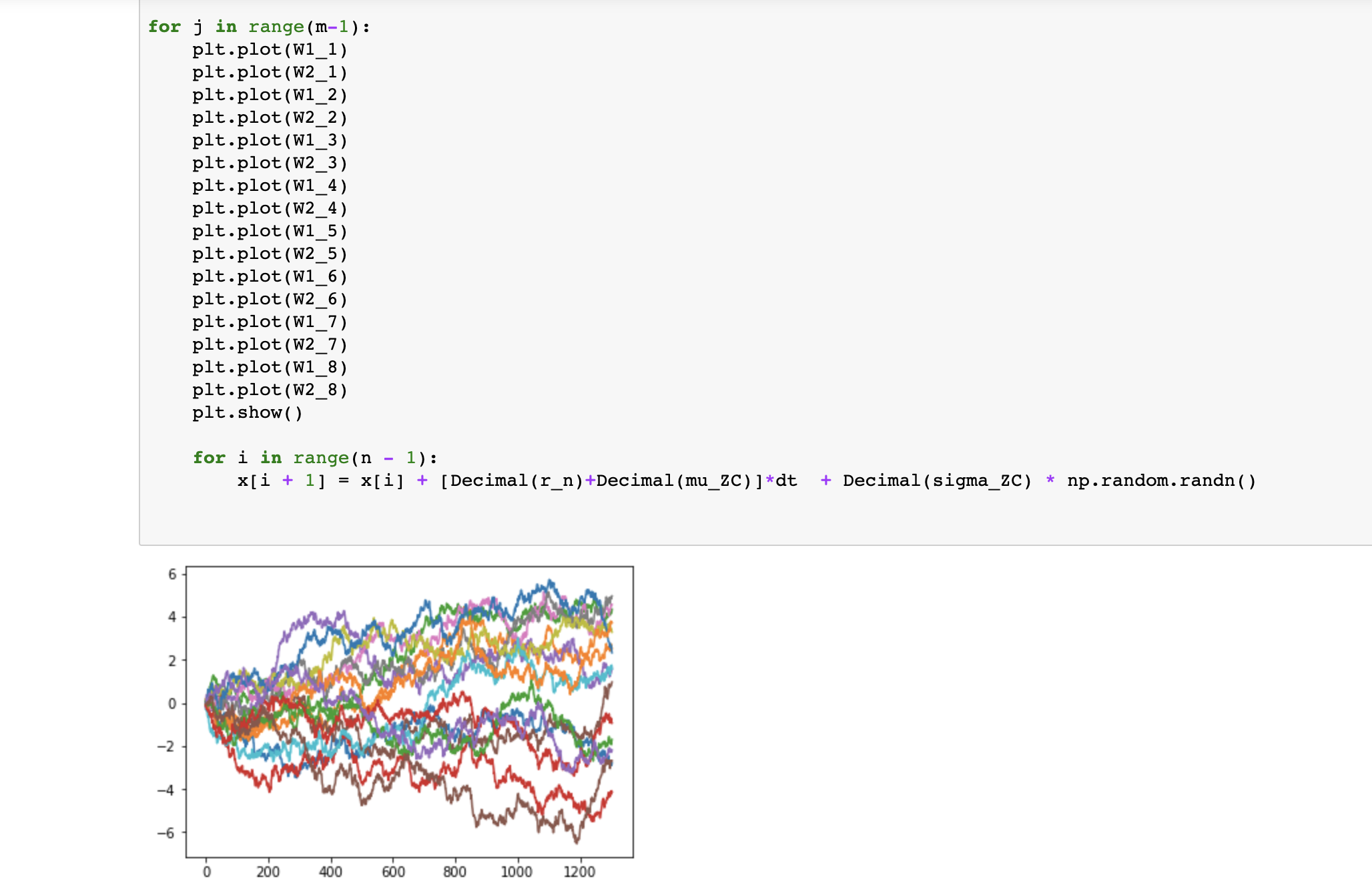}
The simulated price curves for zero-coupon bond and TIPs are shown above.

\subsection{Empirical Model}
For the inflation bond framework, we adopt it from Korn and Kruse model.\\
$r_N(t)$: nominal interest rate for an investment horizon until time t. \\
$r_R(t)$: real interest rate until time t, which is the gain in real purchasing power of an investment. \\
$\mathbb{E}[i(t)]$: expected inflation rate. \\
We have the relation: $r_N(t) = r_R(t) + E[i(t)]$
However, the framework does not illustrate the specific evolution of inflation rate. Hence, I define the inflation rate as:
$dI(r(t)) =  [\theta(t) -a(t)(I(r(t))+i(t))] \ dt + \sigma_1(t) \ dW_1(t)$ \\
where I denotes the inflation rate. \\
I(t) denotes the pricing Index (CPI).\\
Since in general I(t) has positive association(different than correlation) with the inflation index, so we use $I(r(t))+i(t)$ to indicate this positive association. \\
where u has the process below with initial value of 0:
$du = - bu \ dt + \sigma_2 dW_2(t)$ \\
My model is similar to Hull-White model, but with modifications.

\subsection{Inflation}
Digging into the foundation of the pricing model of inflation derivatives, the core essential factor is price index, which determines the inflation rate of the country distributing the financial instruments. \\
As mentioned by Dam\&Macrina\&Skovmand\&Sloth (page 44 in the paper " Rational Models for Inflation-Linked Derivatives):" When inflation-linked pricing is considered, there is an additional layer of modelling complexity that needs to be taken care of: the link between the so-called real and nominal economies, thus the construction of a stochastic model for the (consumer) price index."(In this model, CPI acts like an exchange rate from the nominal to the real economy, due to $C_t = \frac{\pi^R_t}{\pi^N_t}$) \\

Meanwhile, as mentioned by Jarrow \& Yildirim \cite{Robert}, the inflation index is given as: $\frac{dI(t)}{I(t)} = \mu_I(t)dt + \sigma_I(t)dW_I(t)$ with $\mu_I(t)$ random factor and $\sigma_I(t)$ deterministic function. Thus, there is no (consumer) price index involved in the differential equation.\\ 
In Manning \& Jones' model, the index is inferred as a regression process by using the historical data of index: $F^i(t;T_1,T_2) = \frac{I(t;T_2)}{I(t;T_1)} - 1 $. \\
So far, none of the models use stochastic differential equations as approach to the problem. As the first attempt, I define the price index as: \\
\noindent
$\frac{dI(t)}{I(t)} = \mu_I(t)dt + \sigma_I(t)dW_I(t)$ \\
$\frac{d\mu_I(t)}{\mu_I(t)} = \frac{C_{T_2}}{C_{T_1}}dt +\sigma_{\mu}dW_{\mu}(t)$

The rate of inflation realized between $T_1$ and $T_2$ is adopted from the models developed by Dodgson-Kainth \cite{Dodgson}, Bezooyen-et-al and Jarrow-Yildirim: $R^I(T_1,T_2)= \frac{I(T_2)}{I(T_2)} -1 $ \\

\subsection{Conclusion}
In the general form of SDE, e.g. $dX_T = X_t (\mu dt + \sigma dW)$ indicates that the "small segement" of difference between $X_T$ and $X_t$ is determined by: \\
1. the historical information of $X_t$ at only time t.
2. the rate $\mu$ of increasing or decreasing in the segment of time $dt$, 

\subsection{Analytical solution of SDE}
If the drift term $\mu(.)$ and diffusion term $\sigma(.)$ in a SDE satisfies the conditons below:\\
1. Global Lipschitz Condition: For all $x,y \in \mathbb{R}$ and $t \in [0,T]$, there exists a constant $K < + \infty$ to let: \\
$|\mu(t,x)- \mu(t,y)| + |\sigma(t,x) - \sigma(t,y)| < K|x - y|$ \\
2. Linear Growth Condition: For all $x \in \mathbb{R}$ and $t \in [0,T]$, there exists a constant $C < + \infty$ to let: \\
$|\mu(t,x)| + |\sigma(t,x)| < C(1 + |x|)$ \\ 
Then there exists an unique and continuous strong solution to let:  \\
$\mathbb{E}(\int^T_0 |X_t|^2 \ dt) < \infty$ \\
For example, the solutions for Brownian motion and Geometric Brownian motion are given below: \\
Brownian Motion: \\
$dX_t = \mu \ dt + \sigma \ dW_t$ \\
With given initial condition value $X_{t_0}$, the solution of this SDE is:\\
$X_t = X_{t_0} + \mu(t-t_0) + \sigma(W_t - W_{t_0})$\\
Geometric Brownian Motion:\\
$dX_t = \mu X_t \ dt + \sigma X_t \ dW_t$\\
Given initial value condition $X_{t0}$, the solution of the SDE is:
$X_t = X_{t_0} exp[(\mu - \frac{\sigma^2}{2})(t-t_0) + \sigma(W_t - W_{t_0})]$

\subsection{Ito Lemma}
Suppose X(t) satisfies SDE:\\
$dX_t = \mu(t,X_t) \ dt + \sigma(t,X_t) \ dW_t$ \\
Since $f(X)$ is the function of X, then:\\
$df(X) = f_x(X) \ dX + \frac{1}{2}f_xx(dX)^2$\\
The expansion of $(dX)^2$ obeys the following rule:\\
\includegraphics{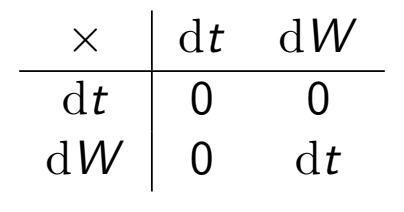}

\subsection{The application of Ito Lemma}
Suppose $X(t)$ satisfies SDE: \\
$dX_t = \mu X_t \ dt + \sigma X_t \ dW_t$ \\
In order to solve the solution of $X(t)$, suppose $Y(t) = logX(t)$, then $\frac{\partial Y}{\partial X} = \frac{1}{X}$, $\frac{\partial ^2 Y}{\partial X^2} = -\frac{1}{X^2}$, According to Ito lemma: \\
$dY = \frac{\partial Y}{\partial X} \ dX + \frac{1}{2} \frac{\partial ^2 Y}{\partial X^2}(dX_t)^2 \\
= \frac{1}{X}(\mu X dt + \sigma X dW) + \frac{1}{2}(- \frac{1}{X^2}) \sigma ^2 X^2 dt\\
= \mu dt + \sigma dW - \frac{1}{2} \sigma ^2 dt \\
= (\mu - \frac{1}{2} \sigma ^2) \ dt +\sigma dW$ \\
Then $Y(t)$ is Brownian motion: \\
$Y(t) = Y(t_0) + (\mu - \frac{1}{2}\sigma^2)(t-t_0) + \sigma (W(t) - W(t_0))$\\
$X(t) = exp(Y(t))$\\
$=X(t_0)exp[(\mu - \frac{1}{2} \sigma^2)(t-t_0) + \sigma(W(t)-W(t_0))]$

\subsection{Numerical Solution of SDE}
Not all stochastic differential equations are capable for discovering an analytical solution. Another approach is to find a numerical solution for SDE. There are few catalogs of numerical methods, such as Monte Carlo simulation, tree method, and traditional numerical methods. 
\subsection{Traditional Numerical Methods}
The two typical numerical methods for sovling SDE are Euler method and Milstein method. \\
Euler: \\
$Y_{i+1} = Y_{i} + \mu(t_i,Y_i)(t_{i+1}-t_i) + \sigma(t_i,Y_i)(W_{i+1}-W_i)$ \\
Milstein: \\
$Y_{i+1} = Y_{i} + \mu(t_i,Y_i)(t_{i+1}-t_i) + \sigma(t_i,Y_i)(W_{i+1}-W_i) + \frac{1}{2}\sigma(t_i,Y_i)\sigma_x(t_i,Y_i)[(W_{i+1}-W_i)^2-(t_{i+1}-t_i)]$ \\ 
where $\sigma_x$ denotes the derivative of $\sigma$ with respect to $x$. In addition, $W_{i+1}-W_i = \sqrt{t_{i+1} - t_i}Z_{i+1}, i = 0,...,n-1$, $Z_1,...,Z_n$ are normally distributed and independent from each other.

\subsection{Monte Carlo Simulation}
Monte-Carlo simulation is a standard and widely used numerical method. The ideology is to simulate the possible paths of the expectation value that we are seeking for, then average all the simulated values to calculate the result. The advantage of Monte-Carlo simulation comes to its generality. For example ,the models of exotic options have jump processes(which can be exercised before maturity), which involve a lot of computations, regression-based Monte-Carlo method can simplify the computation and be applied to solve pricing problem of exotic option.\\
The disadvantage of Monte-Carlo Simulation is time complexity. In order to reduce the time complexity, we can use multi-level Monte-Carlo, use variance reduction, or run on a super computer.
\subsection{Lattice Based method}
The tree method includes binary(binomial) tree and ternary(trinomial) tree.
With known of volatility $\sigma$, a n-level binary can be constructed by the formulas(as up-movement and down-movement): \\
$u=exp(\sigma \sqrt{T/N})$ \\
$d=exp(-\sigma \sqrt{T/N})$ \\
In the case of backwardation, the initial value of an option can be calculated by transversing the tree.\\
In the case of forward rate model, for example Ho-Lee model given by $dr_t=\theta_t \ dt + \sigma \ dW_t$, the calculation can be done as following: \\
Applying Euler method, the differential equation becomes: \\
$r_t = r_{t - \Delta t} + \theta_{t- \delta t} \delta t +\sigma \sqrt{\Delta t} Z $
where Z is a standard normal random variable. \\
Under the property of discrete time, we then have \\
$r_t  = r_0 + \Delta t \sum_{t_0 \leq t_i \leq t - \Delta t} \theta_{t_i} + \sigma \Delta t \sum_{t_0 \leq t_i t-\delta t} Z$ \\
Then the price of zero-coupon bond under no arbitrage condition involves from:\\
$P(0,T)=\mathbb{E}^Q[exp(- \int^T_0 r(t)ds)]$ \\
into \\
$P(0,t_n) = \mathbb{E}^Q[exp(-\delta t \sum^{n-1}_{i=0}r(t_i))]$\\
where $T=t_n$ and $\mathbb{E}^Q$ denotes the expectation under Q-measure \\
For example, at time $n=2$, the equation becomes: \\
$P(0,t_2) = \mathbb{E}^Q[\Delta t * exp(-r_{t_0}-r_{t_1})]=exp(-\Delta t* r_{t_0})\mathbb{E}^Q[exp(-\Delta t * r_{t_1})]$ \\
In terms of Variation: \\
$P(0,t_2) = exp(-\Delta t r_{t_0}) exp(-\Delta t \mathbb{E}^Q[r_{t_1}] + \frac{1}{2} \Delta t Var^Q[r_{t_1}])$ \\
Since $r_t$ is normally distributed: \\
$lnP(0,t_2)=-\Delta t r_{t_0} -\Delta t r_{t_0} - \Delta t \theta_0 +\frac{1}{2}\sigma^2(\Delta)^2 = -2 \Delta t r_{t_0} - \theta_0 \delta t +\frac{1}{2} \sigma^2(\delta)^2$ \\
Meanwhile : $lnP(0,t_2) = \Delta t [-lnP(0,0) - lnP(0,t_1)]$\\
Thus, $-r_{t_0}-lnP(0,t_1)= -2r_{t_0} - \theta_0 t +\frac{1}{2}\sigma^2 \Delta$ \\
So we have $\theta_{t_0} = lnP(0,t_1)-r_{t_0}+\frac{1}{2}\sigma^2 \Delta t$\\
With this updated $theta$ for each $t_n$, we are able to calculate the forward interest curve.\\

As for ternary tree, the only ternary tree method working under the assumption of complete market is: \\
$min[\frac{1+R-d}{l-d},\frac{u-(1+R)}{u-l}]$ \\
$u: up-movement$ \\
$l: mid-movement$ \\
$d: down-movement$ \\
Sometimes, the backward stochastic differential eqation(BSDE) is invloved in seeking for the solution of a pricing model:\\
$https://www.mathematik.hu-berlin.de/~perkowsk/files/bsde.pdf$
\subsection{PDE method}
$https://www.math.nyu.edu/faculty/goodman/teaching/StochCalc2018/notes/Lesson5.pdf$
Beside approching to the solution of SDE directly by analytical or numerical method, converting SDEs into PDEs is also commonly adapted. 
The generalized method of converting a SDE into a PDE is by applying Feymann-Kac theorem. 
Converting the SDE into Feymann-Kac PDE reveals a connection. Intrinsically, the PDE converted by applying Feymann-Kac is the Kolmogorov forward equation of its original SDE. Moreover, a single SDE has two different kinds of operators: forward operator and backward operator, which correlate with the drift term and volatility term. In general, forward PDE with forward operator is subject to the initial value problem, seeking for the solution of final status; backward PDE with backward operator depends on the boundary condition at final value(terminal condition), seeking for the solution of initial status. 
There is a $Green's function$, or so called $transition density$

\subsection{Characteristic Function}
For the models with Levy process, which is an adapted stochastic process started from value zero at time zero, with stationary and independent increments, we can apply Levy-Khintchine formula and seek for its solution. \\

Levy-Khintchine: \\

$\Psi(\lambda)$ is the characteristic function of an infinitely divisible distribution if and only if: \\

$\Psi(\lambda)=i\langle a,\lambda\rangle +\frac12 Q(\lambda)+\int_{\mathbb{R}^d}(1-e^{i\langle \lambda,x\rangle}+i\langle \lambda,x\rangle 1_{|x|<1})\Pi(dx).$

for $a\in\mathbb{R}^d$, $Q$ a quadratic form on $\mathbb{R}^d$, and $\Pi$ a so-called Levy measure satisfying $\int (1\wedge |x|^2)\Pi(dx)<\infty$.

$i\langle a,\lambda\rangle$ comes from a drift of -a. Note that a deterministic linear function is a (not especially interesting) Levy process.
$\frac12Q(\lambda)$ comes from a Brownian part $\sqrt{Q}B_t$.
The rest corresponds to the jump part of the process.
$https://eventuallyalmosteverywhere.wordpress.com/2012/12/04/the-levy-khintchine-formula/$

\section{Math knowledge}
\subsection{The Review of Fundamental Knowledge}
The two-variable function $X(t,\omega)$ is a stochastic process, $\omega \in \Omega$ ($\Omega$ is a sample space). If $t \in \mathbb{N}$, then $X(t,\omega)$ is a discrete time process. If $t \in \mathbb{R}^+$, then $X(t,\omega)$ is a continuous time process, written as $X={X_t, t \leq 0}.$ Sometimes, we use X(t) to represent $X_t$.

For a fixed $\omega$, $X(.,\omega): T \rightarrow \mathbb{R}$ is a sample path(trajectory) for each $\omega \in \Omega.$

For a fixed t, $X(t,.): \Omega \rightarrow \mathbb{R}$ is a random variable for each $t \in T$.

A Gaussian process is a stochastic process for which any joint distribution is Gaussian. A stochastic process is strictly stationary if it is invariant under time displacement and it is wide-sense stationary if there exist a constant $\mu$ and a function $c$ such that

$\mu_t = \mu, \quad \sigma_t^2 = c(0), \quad C_{s,t} = c(t-s) \qquad \forall s,t \in T$

A stochastic process is a martingale if 

$E(X_t | \mathbb{A}_s = X_s) \qquad \forall 0 \leq s \leq t$

\subsubsection{Brownian Motion/The Wiener Process}
Brownian motion(also called Wiener process) is the random motion of particles suspended in a fluid resulting from their collision with the fast-moving molecules in the fluid. Brownian motion was discovered by a Scotland biologist, Robert Brown, under the observation of pollen in fluid by a microscope.

Math property: 

Suppose the continuous time process $W_t: 0 \leq t < T$ is a standard Brownian motion on $[0,T)$ where $W_0=0$.

Independence: For $0 \leq t_1 < t_2 < ... < t_n <T$, the stochastic processes $W_{t_2} - W_{t_1}, W_{t_3} - W_{t_2}, ..., W_{t_n} - W_{t_{n-1}}$ are independent.

Normal distribution: For any $0 \leq s < t < T$, $W_t - W_s$ obeys the normal distribution with average 0 and standard deviation $t-s$. 

\subsubsection{Stochastic Process}
In probability theory and related fields, a stochastic or random process is a mathematical object usually defined as a family of random variables.\cite{stochastic} The random variables are defined on a common probability space $(\Omega,\mathcal{F},P)$, where $\Omega$ is a sample space, $\mathcal{F}$ is a $\sigma$-algebra, and $P$ is a probability measure and the random variables which indexed by some set $T$ and all take values in the same mathematical space $S$ must be measurable with respect to some $\sigma$-algebra $\Sigma$.\cite{John}

In other words, for a given probability space $(\Omega,\mathcal{F},P)$ and a measurable space $(S,\Sigma)$, a stochastic process is a collection of $S$-valued random variables, which can be written as: $ \{ X(t):t \in T \} $.\cite{Ionut} Every stochastic process can be viewed as a function of two variables $-t$ and $\omega$. For each fixed $t$, $\omega \rightarrow X_t(\omega)$ is a random variable, as postulated in the definition.\cite{stochastic2}

\subsubsection{Markov Process}
Markov processes are stochastic processes, traditionally in discrete or continuous time, that have the Markov property, which means the next value of the Markov process depends on the current value, but it is conditionally independent of the previous values of the stochastic process.\cite{Markov4} In other words, the behavior of the process in the future is stochastically independent of its behavior in the past, given the current state of the process.\cite{Markov3} The Markov process $X_t$ is ergodic if the time average on [0,T] for $T \rightarrow \infty$ of any function $f(X_t)$ is equal to its space average with respect to (one of) its stationary probability densities.\cite{markov2} In mathematical language, we denote Markov Process by: a stochastic process $(X_t)_t \in I$ on $(\Omega,\mathcal{U},P)$ with state space $(S,\mathcal{B})$ is called an $(\mathcal{F}_t)$ Markov process iff $(X_t)$ is adapted w.r.t the filtration $(\mathcal{F}_t)_t \in I$, and\cite{markov}
\begin{equation}
P[X_t \in B|\mathcal{F}_s] = P[X_t \in B|\mathcal{X}_s]
\qquad
\end{equation}
where $P$-a.s. for any $B \in \mathcal{B}$ and $s,t \in I$ with $s \leq t$.

Theorem:

The solution $x(t,\omega)$ of a stochastic differential equation with $x(0,\omega)=x_0(\omega)$ is a Markov process on the interval $[0,T]$ with the initial distribution\cite{markov5}
\begin{equation}
P \{ x(0,\omega) \in A \} =P_0 \{ A \}
\end{equation}
and the transition probabilities given by
\begin{equation}
P \{ s,x,t,A \} =P \{ x(t,\omega) \in A|x(s,\omega)=x \}
\end{equation}
for all $0 \leq s \leq t \leq T$.\cite{Markov6}
\subsubsection{Diffusion process}
In probability theory and statistics, a diffusion process is a solution to a stochastic differential equation. It is a continuous-time Markov process with almost surely continuous sample paths. Brownian motion, reflected Brownian motion and Ornstein–Uhlenbeck processes are examples of diffusion processes.\cite{diffusion}
\subsubsection{Ito Integration}
Suppose ${X(t), 0 \leq t \leq T}$ is a stochastic process based on Brownian motion, and satisfies:
\begin{equation}
\int^T_0 \mathbb{E}(X(s)^2) \ ds < + \infty
\qquad
\end{equation}

Then the Ito integration of X is defined as:
\begin{equation}
I_t(X) = \int^t_0 X_s \ dW_s = \lim_{\Vert \Pi_n \Vert \rightarrow 0} \sum^{n-1}_{i=0} X(t_i)(W(t_{i+1})-W(t_i)) 
\qquad
\end{equation}

For example, 
\begin{equation}
\int^t_{t_0} dW(s) = lim_{n \rightarrow \infty} \sum(W(t_{i+1})-W(t_i)) = W(t) - W(t_0)
\qquad
\end{equation}
\subsubsection{Ito Process}
The stochastic process ${X(t), 0 \leq t \leq T}$ is written as:
\begin{equation}
X_t = X_0 + \int^t_0 g(s) \ ds + \int^t_0 h(s) \ dW_s
\qquad
\end{equation}

In addition, $g(t,\omega)$ and $h(t, \omega)$ are two different adapted processes, and satisfy:
\begin{equation}
P{\int^T_0 |g(t,\omega)| \ dt < \infty} = 1, \quad {\int^T_0 h(t,\omega)^2 \ dt < \infty} = 1
\qquad
\end{equation}
then ${X(t), 0 \leq t \leq T}$ is a Ito process. 
\begin{equation}
X_t = X_0 + \int^t_0 \mu(s,X_s) \ ds + \int^t_0 \sigma(s,X_s) \ dW_s
\qquad
\end{equation}
Written in form of differential terms: 
\begin{equation}
dX_t = \mu(t,X_t) \ dt + \sigma(t,X_t) \ dW_t
\qquad
\end{equation}
Sometimes, in a concise form: 
\begin{equation}
dX_t = \mu(X_t) \ dt  + \sigma(X_t) \ dW_t
\qquad
\end{equation}
$\mu(.)$ is so called drift term, and $\sigma(.)$ is diffusion term.

There are few representative processes, such as geometric Brownian motion, Ornstein-Uhlenbeck process (Vasicek process), Cox-Ingersoll-Ross process. 

Geometric Brownian Motion: 
\begin{equation}
dX_t = \mu X_t \ dt + \sigma X_t \ dW_t
\qquad
\end{equation}

Ornstein-Uhlenbeck Process(Vasicek Process):
\begin{equation}
dX_t = (\theta_1 - \theta_2 X_t) \ dt + \theta_3 \ dW_t
\qquad
\end{equation}
When $\theta_2 > 0$, this process has the property of mean-reverting, and can be expressed as: 
\begin{equation}
dX_t = \theta(\mu - X_t) \ dt + \sigma \ dW_t
\qquad
\end{equation}
where $W_t$ is a Wiener process under the risk neutral framework modelling;

\noindent$\mu$: "long term mean level". All future trajectories of $X_t$ will evolve around a mean level $\mu$ in the long run;

\noindent$\theta$: "speed of reversion". $\theta$ characterizes the velocity at which such trajectories will regroup around $\mu$ in time;

\noindent$\sigma$: "instantaneous volatility", $\sigma$ measures the instant by the quantity of randomness $\frac{\sigma^2}{2 \theta}$.

Cox-Ingersoll-Ross Process:
\begin{equation}
dX_t = (\theta_1 - \theta_2 X_t) \ dt + \theta_3 \sqrt{X_t} \ dW_t
\qquad
\end{equation}
When $2 \theta_1 > \theta_3^2$, this process is strictly positive, can also be written as:
\begin{equation}
 dX_t = \theta(\beta - X_t) \ dt + \sigma \sqrt{X_t} \ dW_t
\qquad
\end{equation}

The Cox-Ingersoll-Ross (CIR) model determines the interest rate as a product of current volatility, the mean rate and spreads. Since there is a square root term in the equation, the model indicates the mean reversion towards a long-term normal interest rate level. In summary, the CIR is a one-factor equilibrium model utilizing a square-root diffusion process mainly for forecasting interest rates. The similarity of CIR model and Vasicek model is that they are both one-factor model. The difference is that Vasicek model does not include a square-root component, which allows the growth rate to be negative.
 
\subsection{Math tools}
\subsubsection{Markov Property}
The Markov property states as: "Given the present state, $B_s$ , any other information about what happened before time $s$ is irrelevant for predicting what happens after time $s$." Since the Brownian motion is translation invariant, the Markov property can be simplified as: "if $s \geq 0 $ then $B_{t+s} - B_s, t \geq 0$ is a Brownian motion that is indepedent of what happened before time s." This implies that if $s_1 \leq s_2 ... \leq s_m$ and $0 \leq t_1 ... \leq t_n$, then $(B_{t_1 + s} - B_s, ..., B_{t_n+s}-B_s)$ is independent of $(B_{s_1},...,B_{s_n})$. 

\subsubsection{Ito Lemma}
In mathematics, Ito's lemma is an identity used in Ito calculus to find the differential of a time-dependent function of a stochastic process. It serves as the stochastic calculus counterpart of the chain rule. It can be heuristically derived by forming the Taylor series expansion of the function up to its second derivatives and retaining terms up to first order in the time increment and second order in the Wiener process increment. The lemma is widely employed in mathematical finance, and its best known application is in the derivation of the Black–Scholes equation for option values.\cite{Itolemma}

Lemma:

The following equation holds for Brownian motion $W_t$ and quadratic derivative functions$f(x)$:

\begin{equation}
df(W_t)=f'(W_t) \ dW_t+\frac{1}{2}f''(W_t) \ dt
\end{equation}

The following equation holds for Brownian motion $X_t$ and quadratic derivative functions$f(t,x)$:

\begin{equation}
df(t,X_t)=(\frac{\partial f}{\partial t}+\frac{1}{2} \frac{\partial^2 f}{\partial x^2}) \ d<X,X>_t+\frac{\partial f}{\partial x} \ dX_t
\end{equation}

Define the Ito process as a stochastic process that satisfies the following stochastic differential equations:

\begin{equation}
dX_t=\mu_t \ dt+\sigma_t \ dW_t
\end{equation}

The following equation holds for Ito process $X_t$ and quadratic derivative functions$f(t,x)$:

\begin{equation}
df(t,X_t)=(\frac{\partial f}{\partial t}+\mu_t\frac{\partial f}{\partial x}+\frac{1}{2} \sigma_t^2\frac{\partial^2 f}{\partial x^2}) \ dt+\sigma_t\frac{\partial f}{\partial x} \ dW_t
\end{equation}

\subsubsection{laplace transformation}
The Laplace transform is an integral transform. It transforms a function of a real variable $t$ to a function of a complex variable $s$. \cite{Doetsch} The Laplace transform of a function $f(t)$, defined for all real numbers $t \ge 0$, is the function $F(s)$, which is a unilateral transform defined by
\begin{equation}
F(s)=\int_{0}^{\infty}{f(t)e^{-st} \ dt}
\qquad
\end{equation}
where $s=\sigma+i\omega$ with real numbers $\sigma$ and $\omega$

\subsubsection{Girsanov's theorem(Change of Measure)}
Though the martingale property is not preserved under measure changes, the semi-martingale property remains under measure change. In addition, Girsanov's theorem provides a methodology to sate the precise semi-martingale decomposition under the new measure Q.\cite{girsanov}

We begin with a probability space$(\Omega,\mathcal{F},\mathbb{P})$ and a non-negative random variable $Z$ satisfying $\mathbb{E}Z=1$. We then define a new probability measure $\widetilde{P}$ by the formula 
\begin{equation}
 \label{G1}
   \widetilde{P}(A)= \int_AZ(\omega) \ dP(\omega) \qquad \forall A \in \mathcal{F}
\end{equation}

Any random variable X now has two expectations, one under the original probability measure $\mathbb{P}$, which is denoted by $\mathbb{E}X$, and the other under the new probability measure $\widetilde{P}$, which is denoted by $\widetilde{E}X$. These are related by the formula

\begin{equation}
    \widetilde{E}X = \mathbb{E}[XZ]
\end{equation}

If $\mathbb{P}{Z>0} = 1$, then $\mathbb{P}$ and $\widetilde{P}$ agree which sets have probability zero and $\widetilde{E}X = \mathbb{E}[XZ]$ has the companion formula

\begin{equation}
    \mathbb{E}X = \widetilde{E}[\frac{X}{Z}]
\end{equation}

We say $Z$ is the Radon-Nikodymn derivative of $\widetilde{P}$ with respect to $\mathbb{P}$, and we write 
\begin{equation}
    Z = \frac{d\widetilde{P}}{d\mathbb{P}}
\end{equation}
This is supposed to remind us that $Z$ is like a ratio of these two probability measures. In the case of a finite probability model, we actually have 
\begin{equation}
    Z(\omega) = \frac{d\widetilde{P}(\omega)}{d\mathbb{P}(\omega)}
\end{equation}
If we multiply both sides by $\mathbb{P}(\omega)$ and then sum over $\omega$ in a set A, we obtain 
\begin{equation}
    \widetilde{P}(A) = \sum_{\omega \in A}Z(\omega)P(\omega) \qquad \forall \ A \subset \Omega.
\end{equation}

\subsubsection{Girsanov (One dimension)}
Let $W(t)$, $0 \leq t \leq T$, be a Brownian motion on a probability space $(\Omega, \mathcal{F},\mathbb{P})$, and let $\mathcal{F}(t)$,$0 \leq t \leq T$, be a filtration for this Brownian Motion. Let $\Theta(t)$,$0 \leq t \leq T$, be an adapted process and define:
\begin{equation}
    Z(t) = exp[-\int^t_0\Theta(u) \ dW(u) - \frac{1}{2}\int^t_0\theta^2(u) \ du], \quad \widetilde{W}(t) = W(t) + \int^t_0 \Theta(u) \ du,
\end{equation}
and assume that 
\begin{equation}
    \mathbb{E}\int^T_0\theta^2(u)Z^2(u) \ du < \infty 
\end{equation}

Set $Z=Z(T)$, Then $\mathbb{E}Z=1$ and under the probability measure $\mathbb{P}$ given by

\begin{equation}
 \label{G1}
 \widetilde{P}(A)= \int_AZ(\omega) \ dP(\omega) \qquad \forall \ A \in \mathcal{F}.
\end{equation}
The process $\widetilde{W}(t), 0 \leq t \leq T$, is a Brownian motion.

\subsubsection{Feynman-Kac}
The Feynman–Kac formula named after Richard Feynman and Mark Kac, establishes a link between parabolic partial differential equations (PDEs) and stochastic processes. When Mark Kac and Richard Feynman were both on Cornell faculty, Kac attended a lecture of Feynman's and remarked that the two of them were working on the same thing from different directions.\cite{Feynman}

Theorem:

Consider the artial differential equation
\begin{equation}
    \frac{\partial u}{\partial t}(x,t)+\mu(x,t)\frac{\partial u}{\partial x}(x,t)+\frac{1}{2}\sigma^2(x,t)\frac{\partial^2 u}{\partial x^2}(x,t)-V(x,t)u(x,t)+f(x,t)=0
\end{equation}
defined for all $x \in \mathbb{R}$ and $t \in [0,T]$, subject to the terminal condition
\begin{equation}
   u(x,T)=\psi(x)
\end{equation}
where $\mu, \sigma, \psi, V, f$ are known functions, $T$ is a parameter and $u: \mathbb{R} \* [0,T] \rightarrow \mathbb{R}$ is the unknown. The Feynman-Kac formula tells us that the solution can be written as a conditional expectation
\begin{equation}
   u(x,t)=E^Q[\int_{t}{T}{e^{-\int_{t}^{r}{V(X_\tau,\tau \ d \tau)}}f(X_r,r) \ dr}+e^{-\int_{t}^{T}{V(X_\tau,\tau \ d \tau)}}\psi(X_T)|X_t=x]
\end{equation}
under the probability measure $Q$ such that $X$ is an Ito process driven by the equation
\begin{equation}
   dX=\mu(X,t) \ dt+\sigma(X,t) \ dW^Q
\end{equation}
with $W^Q(t)$ is a Wiener rocess (also called Brownian motion) under $Q$, and the initial condition for $X(t)$ is $X(t)=x$.

\bibliographystyle{IEEEtran}
\bibliography{bibliography}

\end{document}